\documentclass[traditabstract]{aa}

\newcommand{\FWHM}{{\rm{FWHM}}}
\newcommand{\FWHMEQ}{{\it{FWHM}}}
\newcommand{\her}{{\emph{Herschel}}}

\usepackage{graphicx}

\usepackage{natbib}

\defcitealias{Fischera2012a}{paper~I}

\begin{document}

\title{Estimating distance, pressure, and dust opacity using submillimeter observations of self-gravitating filaments}
\titlerunning{Submm observations of self-gravitating filaments}

\author{J. Fischera \and P. G. Martin}
\institute{Canadian Institute for Theoretical Astrophysics, University of Toronto, 60 St. George Street, ON M5S3H8, 
	Canada}


\abstract{
	We present a detailed study of the surface brightness profiles of dense filaments in \object{IC 5146}
	using recent \her\ observations done with SPIRE. We describe the profile through
	an equilibrium solution of a self-gravitating isothermal cylinder pressure confined by its surrounding medium.
	In this first analysis we applied a simple modified black
	body function for the emissivity, neglecting any radiative transfer effects.
	Overall we found a good agreement of the observed surface brightness profiles with the model.
	The filaments indicate strong self-gravity with mass line densities $M/l>\sim 0.5 (M/l)_{\rm max}$
	where $(M/l)_{\rm max}$ is the maximum possible mass line density.
	In accordance with the model expectations we found a systematic decrease of the \FWHM{}, a steepening
	of the density profile, and for filaments heated by the interstellar radiation field
	a decrease of the luminosity to mass ratio for higher central column density and mass line density. 	
	We illustrate and discuss the possibility of estimating the distance, external pressure, and dust opacity.
	For a cloud distance $D\sim 500~{\rm pc}$ and a gas temperature of $T_{\rm cyl}=10~{\rm K}$ 
	the model implies an external pressure 
	$p_{\rm ext}/k\sim 2\times 10^4~{\rm K~cm^{-3}}$ and an \emph{effective} dust emission coefficient at $250~\mu{\rm m}$
	given by $\delta\kappa_0^{\rm em} \sim 0.0588~{\rm cm^2~g^{-1}}$ where $\delta$ is the dust-to-gas ratio.
	Given the largest estimate of the distance to the cloud complex, 1~{\rm kpc}, the model yields an upper limit 
	$\delta \kappa_0^{\rm em}\sim 0.12~{\rm cm^2\,g^{-1}}$.
	}

\keywords{Methods: observational, Techniques: photometric, ISM: clouds, ISM: dust, extinction, 
	ISM: structure, ISM: individual objects: \object{IC 5146}, Submillimeter: ISM}

\maketitle

\section{\label{sect_introduction}Introduction}

Dense self-gravitating structures are an essential part of the interstellar medium and are the very first
steps in the star formation process. 
%
Cold self-gravitating structures in the interstellar medium (ISM) can be found in spherical, elliptical or elongated
forms. Cold spherical forms are referred to as Bok Globules \citep{Bok1947}  if they appear rather isolated in the ISM. 
They are, as pointed out by \citet{Curry2000}, physically similar to the so-called condensed cores found in molecular clouds.
A still larger fraction of the cold gas 
seems to be located in filamentary structures which are found to be a common phenomena throughout the ISM. 
They are seen in non star forming clouds such as \object{Polaris} as well as in star forming cloud such as \object{Aquila} 
\citep{Andre2010,Mensch2010,Arzoumanian2011}. 
Like the Bok Globules or cores they are well defined dense structures which make them perfect objects to 
study the cold molecular phase.

As most of the star forming cores are found in the filamentary structures it is discussed whether there exists a strong link
between the two which is important for the star formation process and responsible for the initial mass function of the 
stars (IMF). In a common scenario the condensed structures form hierarchically
out of gravitational instabilities: the filaments out of sheets and the cores out of 
filaments \citep{Schneider1979,Gaida1984,Hanawa1993,Curry2000,Andre2010,Mensch2010}. 


Cold clouds located in the interstellar medium are pressure-confined by the surrounding medium \citep{Ebert1955,Bonnor1956,McCrea1957,Nagasawa1987,
Inutsuka1992,Inutsuka1997,Curry2000,FiegePudritz2000a,Kandori2005,Fischera2008,Fischera2011}. In our last
paper (\citet{Fischera2012a}, \citetalias{Fischera2012a}) we have analyzed in detail the physical properties of pressurized cylinders in direct comparison
with pressurized spheres. 

Filamentary structures in the cloud complexes \object{IC 5146}, \object{Polaris}, and \object{Aquila} show a typical
\FWHM{} $\sim 0.1~{\rm pc}$ but a large variation of the central column density over
three orders of magnitude \citep{Arzoumanian2011}. The filaments in
non star forming regions like \object{Polaris} appear to be smaller than the ones in star forming complexes as \object{Aquila}.
The filaments of \object{IC 5146} cover a physical regime in between the two other samples.
The data overall indicate a trend that filaments are larger in size at higher column density. 

In \citetalias{Fischera2012a} we have argued  that the overall trend 
is probably related to massive elongated structures 
with embedded dense filaments which cannot be described through a simple isothermal cloud model.
However, the physical parameters of elongated structures below a certain 
column density can be well explained by a model of an isothermal self-gravitating cylinder which is
pressurized by the surrounding medium. 

This paper presents a pilot study in which we apply
the physical model outlined in \citetalias{Fischera2012a} to analyze the density
structure.
The filaments chosen for study are in the \object{IC 5146} field
($GLON=94^\circ$, $GLAT=-5^\circ$) studied by \citet{Arzoumanian2011}
as part of the Gould Belt Survey key project \citep{Andre2010} of the
\emph{Herschel Space Observatory}.  \object{IC 5146} is a key \her\
field in revealing the possible `common' properties of filaments in
nearby molecular clouds and therefore it is an important one for which
to develop physical models.  
%
We demonstrate how sub-millimeter observations can be used to derive
an independent measurement of the cloud distance and of the ambient
confining pressure.  We further show that the model can be used to
study the dust emission behavior in the sub-millimeter regime.

In a comparable study \citet{Kandori2005} analyzed the column density profile
of an ensemble of Bok-Globules.
The authors found that a model of a self-gravitating pressurized isothermal sphere can reproduce
the observed column density profiles which they derived using extinction measurements. 
They demonstrated that the model can be used to derive distances and pressures of the surrounding medium.


In Sect.~\ref{sect_observations} we present the observational parameters for the filaments and  the
method used to extract surface brightness profiles. The physical model of the filaments
is described in Sect.~\ref{sect_model}. 
The technique used to derive the physical parameters is described in 
Sect.~\ref{sect_modelprofiles} and the results for the individual filaments are presented in 
Sect.~\ref{sect_results}. We discuss the results 
of the profile fitting and the estimates of the distance and pressure in Sect.~\ref{sect_discussion} where 
we also address systematic uncertainties of the applied method. A short summary is given in Sect.~\ref{sect_summary}.

\section{\label{sect_observations}Observations}

\begin{figure*}[htbp]
	\includegraphics[width=0.49\hsize]{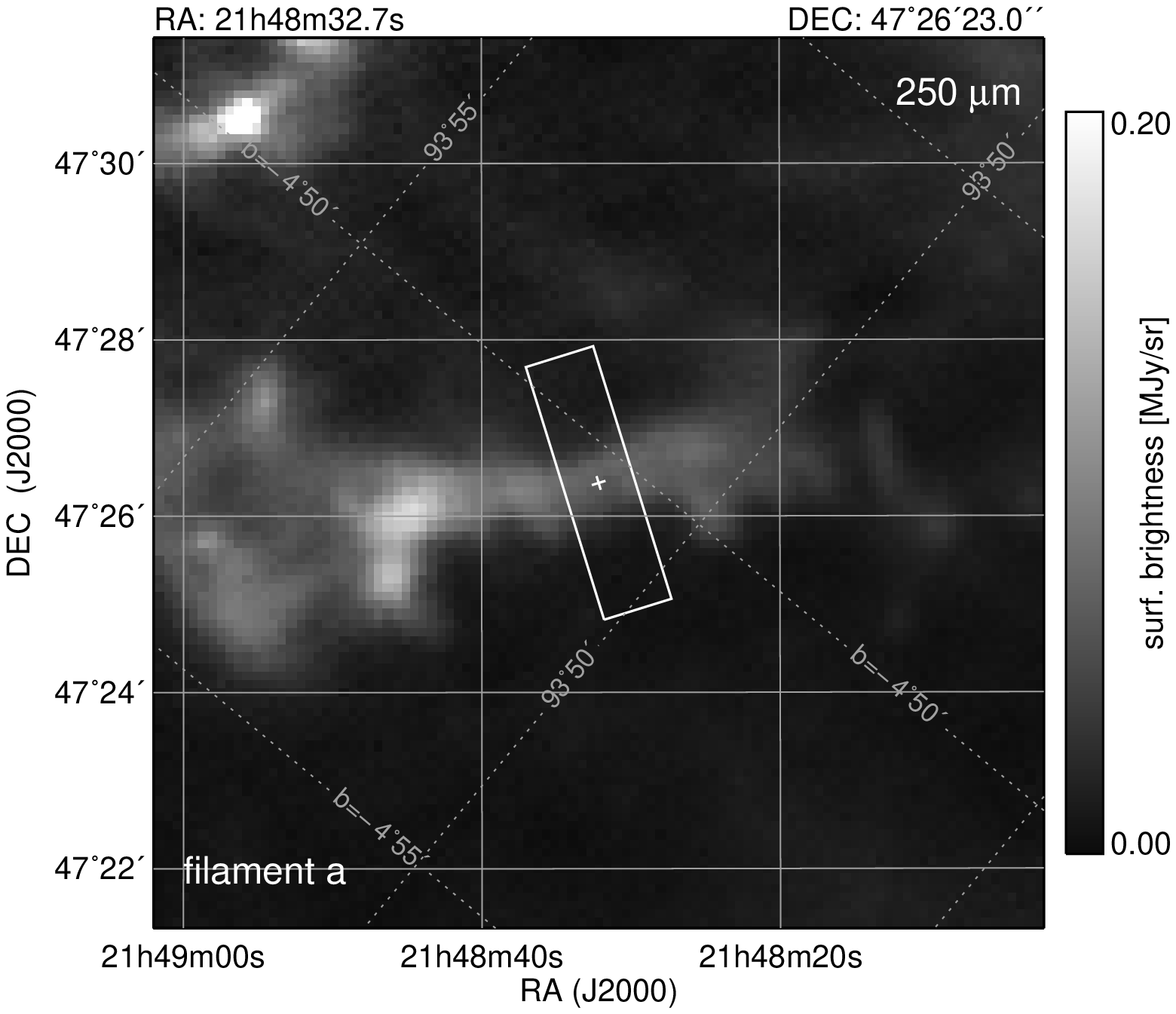}
	\hfill
	\includegraphics[width=0.49\hsize]{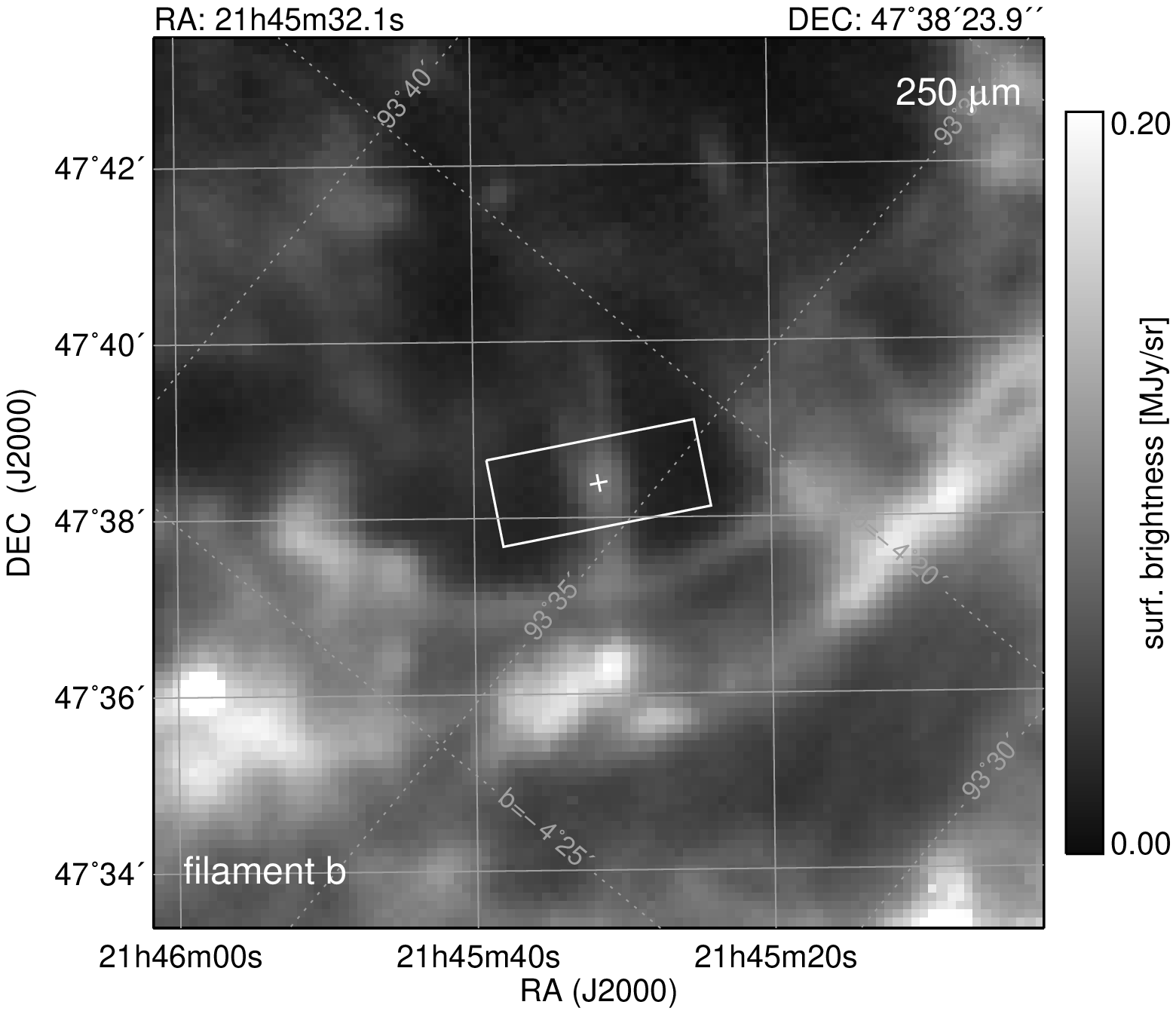}

	\includegraphics[width=0.49\hsize]{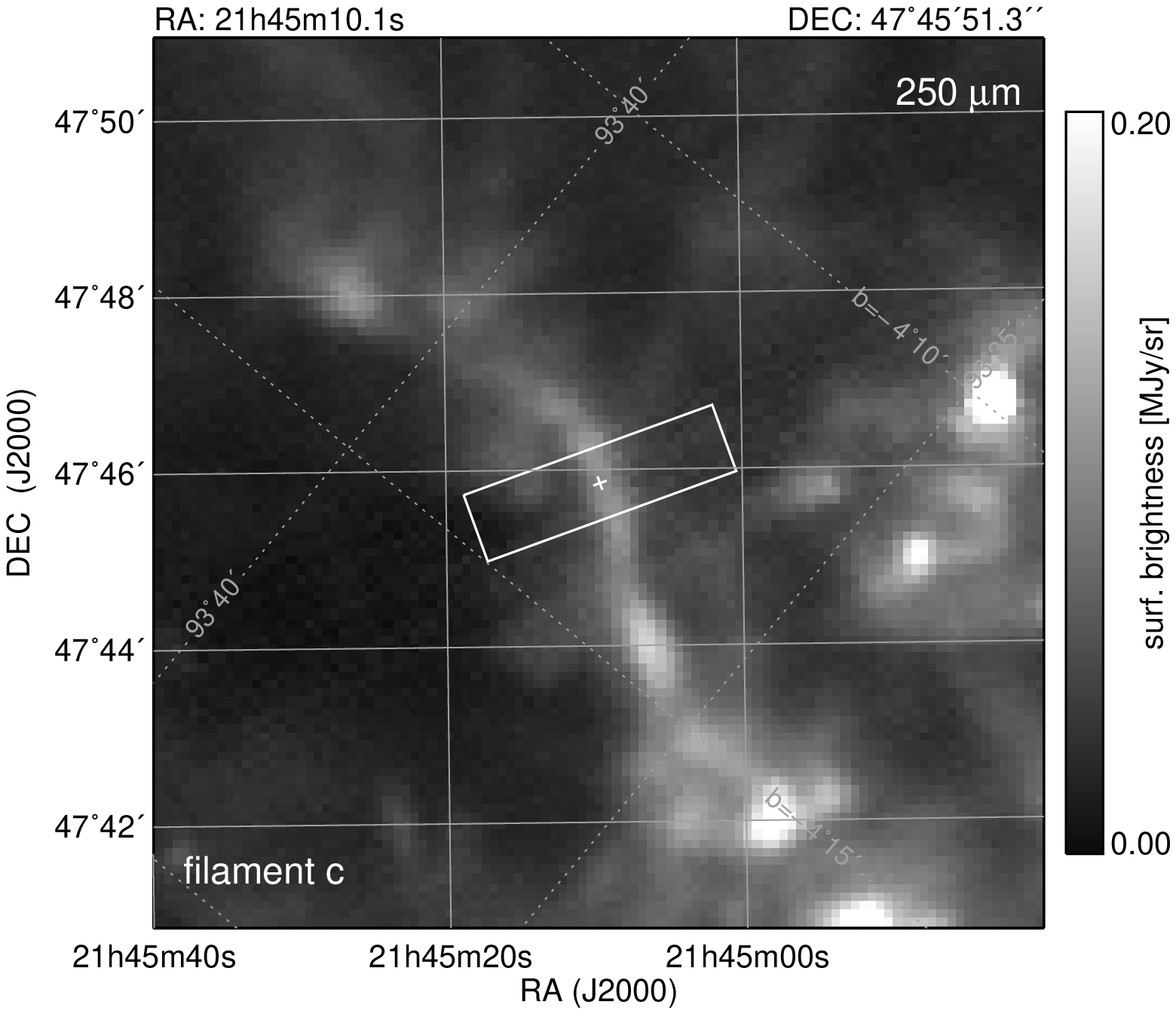}
	\hfill
	\includegraphics[width=0.49\hsize]{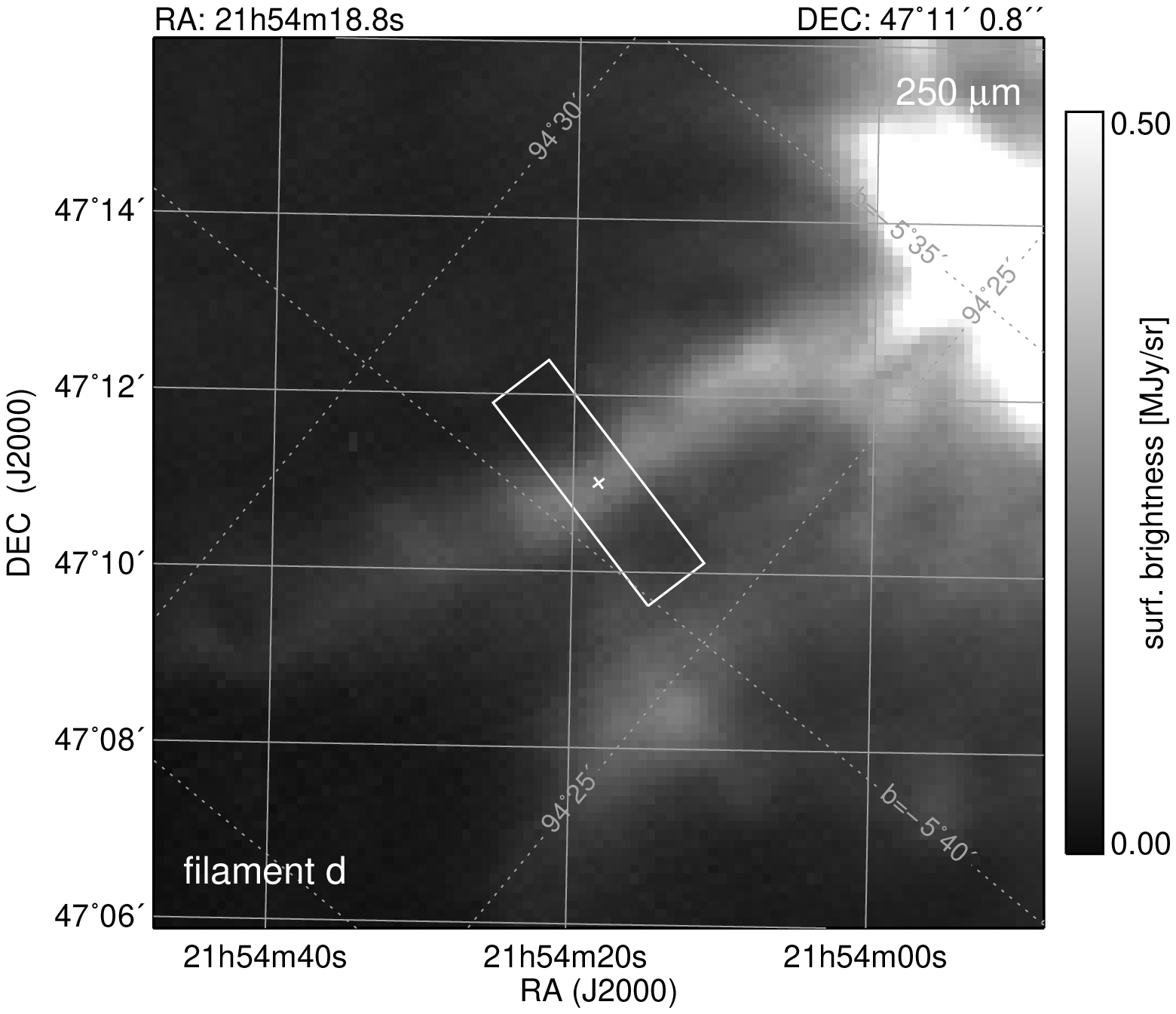}
	\caption{\label{fig_scan}
	SPIRE images taken with SPIRE at reference wavelength 
	$250~\mu{\rm m}$ 
	of four individual regions in \emph{IC 5146} showing the filaments and 
	the apertures used to analyze the surface brightness profiles.}
\end{figure*}

\begin{table*}[htbp]
	\caption{\label{table_coordinates}Scan parameters}
	\begin{tabular}{cccccccc}
		fil. & GLON & GLAT & RA & DEC & $\Theta~[^\circ]$ & length $[']$ & width $[']$ \\
		a	&  $93^\circ 51'10.79''$ & $-4^\circ 50'22.48''$ & 21h48m32.7s & $47^\circ 26'23.0''$ & $+17.23$ &
			3.0 & 0.8 \\
		b & $93^\circ 35' 28.02''$ & $-4^\circ 21' 38.94''$ & 21h45m32.1s & $47^\circ 38' 17.9''$ & $-81.85$ & 
			2.4 & 1.0 \\
		c	& $93^\circ 37' 32.73''$ & $-4^{\circ} 13'29.67''$ & 21h45m10.1s & $47^\circ 45' 51.3''$ & $-70.00$ &
			3.0 & 0.8 \\
		d   	& $94^\circ 27' ~4.40''$ & $-5^\circ39'13.64''$ & 21h54m18.8s	& $47^\circ 11' 0.8'' $ & $+40.51$ &
			2.9 & 0.8 \\
	\end{tabular}
\end{table*}

The study is based on data from the \her\ Science
Archive\footnote{$http://herschel.esac.esa.int/Science_Archive.shtml$}.
The Gould Belt Survey was made in parallel mode, with both PACS
\citep{Poglitsch2010} and SPIRE \citep{Griffin2010} simultaneously,
with two scans of the field in nearly orthogonal directions.  Images
made combining the two scans self-consistently can lead to an optimal
map (a `level 2.5' product), minimizing the striping that appears,
especially for PACS, in images made from a single scan direction (a
`level 2' product).  As we began this study we were able to obtain the
desired `level 2.5' images for SPIRE, but even at the time of writing
(late July 2012) these are not available for PACS data.  We therefore
decided to proceed with SPIRE data only.  

In the future the quality of archival PACS images will certainly
improve (e.g., the promised `Level 2.5' maps combing the cross scans,
plus convergence results from different map-making algorithmns), which
will in principle allow additional constraints on the parmeters of the
model.  However, it is not necessarily simple.  The dust emission
depends more strongly on the temperature gradient in the filament at
shorter wavelengths.  Thus, although the resolution is higher at
shorter wavelengths, the PACS data are more strongly affected by
temperature variations which is actually expected to lead to broader,
not narrower, surface brightness profiles \citep{Fischera2008,
  Fischera2011,Fischera2012a}.  This would make it necessary to use a more complex
model.  The profiles at the submm SPIRE bands are only slightly
affected by temperature variations as the surface brightness in the
Rayleigh-Jeans regime is proportional to dust temperature, not
non-linearly to some higher power.  This makes the SPIRE data used
optimal for analyzing the structure using the simple single dust
temperature model adopted here.

The SPIRE instrument provides photometric images at reference
wavelengths $\lambda_{r}=250$, $350$, and $500~\mu{\rm m}$.  We
demonstrate that the high accuracy of the SPIRE data allow a first
analysis of the density structure using a simplified model and,
despite the incomplete sampling of the spectral energy distribution
(SED), even an estimate of the dust temperature and dust masses. 

The \object{IC 5146} cloud complex has as the most prominent features
a star formation region to the far east called the \object{Cocoon
  Nebula} and a massive elongated structure to the west known as
\object{Northern Streamer}.  The distance to the complex is not well
established with estimates varying by a factor of two from $\sim
460~{\rm pc}$ \citep{Lada1999} to $\sim 1~{\rm kpc}$
\citep{Harvey2008}.  If we assume for the Sun a height
$z_{\odot}=26\pm 3~{\rm pc}$ \citep{Majaess2009} above the mid-plane
of the Galactic disc the position of the cloud complex lies somewhere
between $z=-14~{\rm pc}$ and $z=-61~{\rm pc}$ below the mid-plane.

\citet{Arzoumanian2011} combined SPIRE and PACS observations to
produce column density maps for the same field which were then used to
study the physical properties of individual filamentary structures
including the column density profiles.  In their analysis an automatic
routine was applied to identify individual filamentary structures. The
profiles were derived as an average of numerous orthogonal cuts
through the filaments.

We have chosen a different and a more traditional approach to derive surface brightness profiles by estimating
the profiles on the basis of a rectangular aperture laid over a section of a filament with well defined structure and background.
We have chosen four filaments named `a', `b', `c', and `d' in the paper. The filaments and the apertures
are shown in Fig.~\ref{fig_scan}. The coordinates, sizes, and position angles $\Theta$ for the apertures
are also provided in Table~\ref{table_coordinates}. The orientation and the centre position
of the aperture were chosen to derive an estimate of profiles vertical to the main axis with the mean close to the
centre.

For simplicity for all three filters we have chosen a pixel grid in the aperture with 
pixel size corresponding to that of the image at $250~\mu{\rm m}$ ($6''$). 
The surface brightness at the pixel coordinates of the aperture was derived 
using the IDL cubic interpolation routine (parameter set to -0.5). As the image pixel sizes
at 350 and $500~\mu{\rm m}$ are larger by default ($10''$ and $14''$) the produced 
aperture maps at longer wavelengths are oversampled in respect to the original images.
The surface brightness profiles and uncertainties were derived as the mean value along the aperture
width and its standard deviation.

\subsection{Filament $a$}

Filament $a$ lies in an area located between the \object{Cocoon Nebula} to the west and the \object{Northern Streamer} 
to the east. It has a bone-like structure with a narrow filamentary structure in the centre and wider extensions at 
both ends. The whole filament appears rather isolated with a well defined central part
and an almost perfectly flat background. We have chosen a part with a symmetric surface brightness profile.

\subsection{Filament $b$}

Filament $b$ is located north of the lower main part of the \object{Northern Streamer} which extends from east to west. Interestingly,
filament $b$ lies almost perfectly orthogonal to the main structure. 

\subsection{Filament $c$}

This filament appears as number 5 in the study of \citet{Arzoumanian2011}. It is located north-west from filament $b$ and has a similar
orientation. 
The filament was found to have a large column
density and a rather small \FWHM{} of $\sim 0.06~{\rm pc}$ (for a distant of 460~pc), in fact the smallest in their whole sample.


The aperture contains additional emission in the east unrelated to the filament or the otherwise smooth background.
We modeled and subtracted this emission as described in App.~\ref{app_source}. We restricted the fit by assuming a dust
temperature of $15~{\rm K}$ for this emission;
we further fixed its central position in the direction along the width to be at the boundary of the aperture.

\subsection{Filament $d$}

Filament $d$ is a bright filament which extends to the east of the Cocoon Nebula. 
The bright emission and relatively warm temperature (Sect.~\ref{sect_results}) indicates strong illumination by newly formed stars. 
This is also indicated by the warm dust temperature
of $22.2$~K derived for a filamentary structure located close by to the south-west (filament 18 in the
work of \citet{Arzoumanian2011}).
The radiation field heating the dust grains in the filament might therefore be rather different in spectral shape and strength. 
It is likely also much less isotropic than in the case of the other three filaments. 
The incomplete sampling of the SED is 
more challenging because the temperature is derived on a less strongly curved SED.

\section{\label{sect_model}Model}

The surface brightness profile is in general a result of several
different effects which are related to the intrinsic density profile, dust properties,
illumination of the filament, and in general also temperature variations inside
the structure. 
The method applied
in this paper should be considered as a first approach, rather than a thorough analysis of the
entire problem. In the following we describe the model assumptions.


\subsection{\label{sect_emissivity}Dust emissivity}

%

As mentioned in Sect.~\ref{sect_observations}, the surface brightness
profiles in the SPIRE submm bands are less strongly affected by the
decreasing temperature towards the centre of opaque clouds than are
surface brightness profiles at the shorter wavelength PACS bands.  The
SPIRE observations are quite suitable for tracing the column density
in the context of a single-temperature model, allowing us to ignore in
this pilot study the radial temperature variation inside the cloud.

With this common assumption that the emission from dust grains can be
described by a single dust temperature, the emissivity per gram of
matter (dust and gas) is given by
\begin{equation}
	\label{eq_emissivity}
	\eta_{\nu} =  B_\nu(T_{\rm dust}) \delta \kappa^{\rm em}_0\left(\frac{\nu}{\nu_0}\right)^{\beta^{\rm em}},
\end{equation}
where $B_\nu(T_{\rm dust})$ is the Planck function, $T_{\rm dust}$ the dust temperature, $\kappa^{\rm em}_0$
the \emph{effective} dust emission coefficient at $\lambda_0=c/\nu_0=250~{\mu{\rm m}}$ where $c$ is the velocity of light, 
and $\beta^{\rm em}$ the corresponding \emph{effective} exponent of the frequency dependence. 
$\delta$ is the dust-to-gas ratio in mass for which we assumed $\delta= 0.00588$ as adopted 
from a previous study about the SED of condensed cores \citep{Fischera2011}. 

As the scattering at submm wavelength for interstellar dust grains is negligible the mean extinction is close to the mean absorption
(or emission) properties. For single dust temperatures we would have $\kappa_\lambda^{\rm ext}\sim \kappa_\lambda^{\rm em}$
where the extinction coefficient is given by $\kappa_\lambda^{\rm ext} = \kappa_0^{\rm ext} (\nu/\nu_0)^{\beta^{\rm ext}}$.
In the literature the product $\delta \kappa_\lambda^{\rm em}$ in Eq. ~\ref{eq_emissivity} is therefore replaced by the opacity
$\delta \kappa^{\rm ext}$. 

Here, we explicitly distinguish between the \emph{intrinsic} dust properties which might even vary inside the cloud
and the \emph{effective} ones which correspond to an approximation where the dust emission
peak is modeled by a simple modified black-body function. 

Temperature variations inside the cloud
will for example broaden the SED leading to an apparent lower $\beta^{\rm em}$. As the dust emission is dominated by the warm
dust at the clouds outskirts the single dust temperature assigned to describe the spectral shape is too high and will lead
to an \emph{underestimate} of the dust mass unless the effective dust emission coefficient is reduced
\citep{Fischera2011,Fischera2012b}. The effect can be a factor 2 to 3 depending on the central column density and the
external pressure.

In radiative transfer calculations performed by  \citet{Fischera2008} and \citet{Fischera2011} 
the dust properties were chosen to be close to the 
properties in the diffuse phase with $\kappa_0^{\rm ext}=5.232~{\rm cm^2~g^{-1}}$ and $\beta^{\rm ext} = 2.024$.
For the \emph{effective} emission behavior here we adopted
a constant $\kappa^{\rm em}_0=10 ~{\rm cm^{2}g^{-1}}$ and a spectral index $\beta^{\rm em}=1.8$. 
The higher constant was chosen to compensate for the effect of possible grain growth inside the cloud 
and the lower exponent for possible temperature variations. We will see that the adopted product
$\delta \kappa_0^{\rm em}=0.0588~{\rm cm^2\,g^{-1}}$ is in
broad agreement with the range of distances derived for the cloud complex
(Sect.~\ref{sect_kappa}).


\subsection{Mass surface density}

The filaments are assumed to be isothermal, self-gravitating, and in pressure equilibrium with the ambient medium.
As shown in \citetalias{Fischera2012a} the shape of the density or pressure profile of an infinitely long cylinder can be characterized
through the mass ratio $f_{\rm cyl}=(M/l)/(M/l)_{\rm max}$ given as the ratio of the mass line density $M/l$ 
and the maximum possible
mass line density given by
\begin{equation}
	\label{eq_maxmasslinedensity}
	\left(M/l\right)_{\rm max} = \frac{2K}{G},
\end{equation}
where $G$ is the gravitational constant. The constant $K$ is given by 
$K = k T_{\rm cyl}/(\mu m_{\rm H})$ where $k$, $\mu$, $m_{\rm H}$ are the Boltzmann constant,
the mean molecular weight, and the hydrogen mass. The gas is assumed to be
molecular giving $\mu\approx 2.36$. The effective temperature $T_{\rm cyl}$
contains the thermal and the turbulent motion of the gas. In the case of a non turbulent filament $K=c_{\rm s}^2$ where
$c_{\rm s}$ is the sound speed.

For $T_{\rm cyl}$ we assume $10~{\rm K}$.  As discussed in
\citetalias{Fischera2012a}, this appears to be appropriate generally for dense cold
structures based on molecular line observations.  We also found this
model of pressurized isothermal self-gravitating filaments is in
agreement with observations of the \FWHM{} and central column density
$N_{\rm H}(0)$ of low mass filaments.  This also indicates that the
underlying hypothesis, that these filaments are in hydrodynamical
equilibrium, is self-consistent, which motivates using this model in
the extended analysis in this paper.  It would be ideal to have
diagnostic molecular line observations to examine the temperature and
dynamical state, as indeed there should be in some Gould Belt Survey
fields, but to our knowledge there are none available for the
filaments studied here.  But because $T_{\rm cyl}$ is an explicit
scaling parameter of the model, the systematic dependence of the
results on this choice can be examined (Sect.~\ref{sect_discussion}).
For $10~{\rm K}$, the maximum mass line density is $16.4~M_\odot/{\rm pc}$.

The filament is assumed to be in pressure equilibrium with the surrounding medium
with pressure $p_{\rm ext}$. The mass ratio is related to the overpressure by 
$p_{\rm c}/p_{\rm ext}=1/(1-f_{\rm cyl})^2$ where $p_{\rm c}$ is
the central pressure in the cylinder. 
The mass ratio determines furthermore the shape of the mass surface density while the external pressure its amplitude.
The mass surface density at normalized impact parameter $x=r/r_{\rm cyl}$ where $r_{\rm cyl}$ is the cylinder radius is given by (\citetalias{Fischera2012a})
\begin{eqnarray}
	\label{eq_profile}
	\Sigma_{\rm M}(x) &=& \sqrt{\frac{p_{\rm ext}}{4\pi G}}\frac{\sqrt{8}}{1-f_{\rm cyl}}\frac{1-f_{\rm cyl}}{1-f_{\rm cyl}(1-x^2)}\nonumber\\
			 &&\times \Bigg\{
			 	\sqrt{f_{\rm cyl}(1-f_{\rm cyl})(1-x^2)}\nonumber\\
			&&+\sqrt{\frac{1-f_{\rm cyl}}{1-f_{\rm cyl}(1-x^2)}}\nonumber\\
			&&\times \arctan{\sqrt{\frac{f_{\rm cyl}(1-x^2)}{1-f_{\rm cyl}(1-x^2)}}}\Bigg\}.
\end{eqnarray}
The corresponding column density of H nucleons is given by
\begin{equation}
	N_{\rm H}(x) = \Sigma_{\rm M}(x) (\bar \mu m_{\rm H})^{-1},
\end{equation}
where $\bar \mu \sim 1.4$.

In the model the sky coordinates are related to the impact parameters using
a stretch factor $s = D/r_{\rm cyl}$ so that
\begin{equation}
	s \times (\vartheta-\vartheta_0) = r/r_{\rm cyl} = x,
\end{equation}
where $\vartheta_0$ is the central position of the filament.

As noted in \citetalias{Fischera2012a} the mass surface density profile at given impact parameter
does not depend on $K$ but for given mass ratio and 
external pressure $K$ affects the size of the cylinder. The radius for example is given by
\begin{equation}
	\label{eq_radius}
	r_{\rm cyl} = \frac{\sqrt{8}K}{\sqrt{4\pi G p_{\rm ext}}}\sqrt{f_{\rm cyl}(1-f_{\rm cyl})},
\end{equation}
so that $r_{\rm cyl}\propto T_{\rm cyl}/\sqrt{p_{\rm ext}}$ for given $f_{\rm cyl}$. 
The radius is symmetrical around $f_{\rm cyl}=0.5$ where the cylinder has the maximum physical extension.
The \FWHM{} shows a qualitatively similar behavior as a function of the mass ratio. However,
the maximum value appears at smaller $f_{\rm cyl}$ and at large $f_{\rm cyl}$ the \FWHM{} decreases
as $(1-f_{\rm cyl})$ and is small relative to the physical radius.

\subsection{The SED}

The intrinsic surface brightness of a filament seen at inclination angle $i$ at angle $\vartheta$
is given by
\begin{equation}
	\label{eq_surfbrightness}
	I^{\rm fil}_{\nu}(\vartheta) = \frac{1}{\cos i}\Sigma_{\rm M}(\vartheta) \eta_{\nu}.
\end{equation}

In our sample the profiles are affected by the finite resolution. 
To allow a proper estimate of the intrinsic density structure and the size of the filament we 
convolved the theoretical profile with the corresponding point spread function (psf) 
of the broad band filters. We used for simplicity
the Gaussian approximation where the \FWHM{} of the psf of the three SPIRE filters at
the three reference wavelengths $\lambda_{r}=250~\mu{\rm m}$,  $350~\mu{\rm m}$, and $500~\mu{\rm m}$
is given by $17.6''$, $23.9''$, and $35.2''$, respectively (Table 5.2, SPIRE Observers' Manual, vers. 2.4, June 7, 2011). 

For the assumed model the flux density per unit length is simply given by
\begin{equation}
	\label{eq_flux}
	F^l_\nu = \int{\rm d} \vartheta\,I^{\rm fil}_{\nu}(\vartheta) = \frac{\delta\,M/l}{D\cos i}\,\eta_\nu = \frac{\delta\eta_\nu}{D \cos i} f_{\rm cyl}\frac{2K}{G},
\end{equation}
where $D$ is the distance to the filament.

\begin{table*}[tbp]
	\caption{\label{table_fitmodel}Fitted parameters using the model of a self-gravitating cylinder}
	\begin{tabular}{ccccccc}
	fil. & $\chi^2_{\rm red}$ &$N_{\rm free}$  & $T_{\rm dust}$ & $D\cos i$ & $\log_{10} \frac{p_{\rm ext}/k}{\cos^2 i}$  & $f_{\rm cyl}$ \\
		&		& & [K] & [pc] & [$\log_{10} \rm K~cm^{-3}$]  & \\
	\hline
	\multicolumn{7}{c}{no distance constraint}\\	
	\hline
	a &  0.63 & 79 & $14.8 \pm0.2 $& $493\pm 36 $ & $4.35\pm 0.13$ & $0.68\pm 0.05$ \\	
	b & 0.29 & 61 & $13.9\pm 0.5$ & $377^{+168}_{-325}$ & $5.01_{-0.45}^{+1.24} $ & $0.49_{-0.43}^{+0.19} $  \\
	c & 0.84 & 76 &$13.0 \pm 0.3$ & $488 \pm 31$ & $3.47_{-3.47}^{+0.36}$ & $0.94_{-0.03}^{+0.06}$ \\
	d &  0.16 & 76 & $19.9\pm 0.7$ & $706\pm 70$  & $3.63 \pm 0.24$ & $0.81 \pm 0.05$  \\
		\hline
	\multicolumn{7}{c}{distance constrained ($D=493\pm 36$~pc)}\\	
		\hline
	b & 0.30 & 62 & $14.1\pm 0.4$ & $488 \pm 32$ & $4.71\pm 0.11$ & $ 0.62 \pm 0.07$ \\
	d & 0.24 & 77 & $18.9\pm 0.5$ & $532 \pm 31$ & $4.16 \pm 0.16$ & $0.70 \pm 0.07$ 
	\end{tabular}
\end{table*}
\begin{table*}[tbp]
	\caption{\label{table_fitmodel2}Derived parameters based on Tab.~\ref{table_fitmodel}}
	\begin{tabular}{ccccccc}
	fil. 	& $(M/l)/\cos i$ & $L/M$ &  \FWHMEQ & $\FWHMEQ \cos i$ & $N_{\rm H}(0)/\cos i$ \\
		&  [$M_\odot/{\rm pc}$] & [$L_\odot/M_\odot$] & $[']$ & [0.1 pc] & [$10^{21}{\rm cm^{-2}}$] \\
	\hline
	\multicolumn{6}{c}{no distance constraint}\\	
	\hline
	a	& $11.0\pm 0.8$ & $0.36\pm 0.02$ &  $0.63 \pm 0.03$  & $0.90\pm 0.04 $& $10.4\pm 0.5$ \\
	b	& $8.1_{-7.0}^{+3.0} $  & $0.27 \pm 0.04$ & $0.54 \pm 0.09$ & $0.58_{-0.49}^{+0.15}$ & $12.6 \pm 1.8 $ \\
	c	& $15.4_{-0.5}^{+0.9}$ & $0.18 \pm 0.03$ & $0.35 \pm 0.02 $ & $0.49 \pm 0.05$ & $22.7 \pm 2.4$ \\
	d	& $13.3\pm 0.8$ & $2.18\pm 0.31$ & $0.61 \pm 0.03$ & $1.24\pm 0.11$& $8.5\pm 0.8$ \\
		\hline
	\multicolumn{6}{c}{distance constrained ($D=493\pm 36$~pc)}\\	
		\hline
	b	& $10.1\pm 1.2$ & $0.29\pm 0.03$ & $0.48 \pm 0.03$ & $0.68 \pm 0.05$ &  $12.9 \pm 1.7$\\
	d	& $11.4 \pm 1.1$ & $1.61 \pm 0.18$   & $0.67 \pm 0.03 $ & $1.04\pm 0.06$ & $9.2\pm 0.9$
	\end{tabular}
\end{table*}

An important quantity in the interpretation of the dust emission is obtained by 
integrating over frequency which gives the luminosity to mass ratio \citep{Fischera2011}
\begin{equation}
	\label{eq_lummass}
	L/M = \frac{2\pi D 2 F^l}{(M/l)/\cos i}=\delta\,4 \left<\kappa^{\rm em}_{\rm pl}(T_{\rm dust})\right> \sigma_{\rm SB} T_{\rm dust}^{4},
\end{equation}
where $\sigma_{\rm SB}$ is the Stefan-Boltzmann constant and where
\begin{equation}
	\left<\kappa^{\rm em}_{\rm pl}(T_{\rm dust})\right> = \kappa^{\rm em}_0 \left(\frac{k\lambda_0}{hc}\right)^{\beta^{\rm em}} 
			\frac{\zeta(\alpha)\Gamma(\alpha)}{\zeta(4)\Gamma(4)} T_{\rm dust}^{\beta^{\rm em}},
\end{equation}
is the Planck averaged effective emission coefficient where $\alpha=4+\beta^{\rm em}$ and where $h$ is the Planck constant.
We use this ratio to validate our assumption for the
effective dust emission coefficient (Sect.~\ref{sect_lm}).

\subsection{\label{sect_modelprofiles}Modeling profiles}

The surface brightness profiles for the model filament are determined by the dust temperature $T_{\rm dust}$, 
the external pressure $p_{\rm ext}$, the overpressure $p_{\rm c}/p_{\rm ext}$ or the mass ratio $f_{\rm cyl}$,
the stretch factor $s$, and a position $\vartheta_0$.
It is assumed that the emission at reference frequency $\nu_r$ ($r=1,2,3$)
from the filament is superimposed on a smooth
background of the surface brightness $I_{\nu_{r}}^{bgrd}(\vartheta)$  
modeled through a polynomial function of first (filament a,b, and d) or second order (filament c).

The profiles were modeled using a non-linear $\chi^2$-fit minimizing
\begin{equation}
	\chi^2 = \sum\limits_{r=1}^3 \sum\limits_{i} 
		\frac{(I_{\nu_{r}}(\vartheta_i)-\hat I^{\rm fil}_{\nu_{r}}(\vartheta_i)-I^{\rm bgrd}_{\nu_{r}}(\vartheta_{i}))^2}{\sigma^2_{r,i}},
\end{equation}
where $I_{\nu_{r}}(\vartheta_i)$ is the observed surface brightness at position $\vartheta_i$ 
and $\hat I^{\rm fil}_{\nu_{r}}(\vartheta_i)$ the corresponding theoretical surface brightness from the 
filament  (see App.~\ref{app_surfbrightness}) corrected for spectral shape across the finite bandpass.
In the modeling of the SED we accurately calculated the corresponding theoretical fluxes for the broad band filters
using the appropriate filter transmission curves. 
The color corrections are larger at longer wavelengths where the broad-band fluxes are more than $\sim 10\%$ higher than the corrected values.

The dust temperature itself is sufficiently enough determined through the three broad-band filters. 
Otherwise, there are basically two main parameters determining the fit to the profiles: the mass ratio $f_{\rm cyl}$
for the shape of the profile and the pressure which determines for given $f_{\rm cyl}$ the amplitude. From the
modeling standpoint we found that it is advantageous to perform the modeling in the parameter plane spanned
by the central mass surface density $\Sigma_{\rm M}(0)$ (from Eq.~\ref{eq_profile}) and the mass ratio $f_{\rm cyl}$.

\section{\label{sect_results}Results}

\begin{table}[htbp]
	\caption{\label{table_fluxes}Flux $\nu_r \bar F_{\nu_r}^{l}$ $[10^{-10}{\rm W/m^2}]$}
	\begin{tabular}{ccccc}
		fil. &  $250\mu{\rm m}$ & $350\mu{\rm m}$ & $500\mu{\rm m}$  \\
		\hline
		\multicolumn{4}{c}{no distance constraint}\\
		a & $1.688 \pm 0.023$ & $0.790 \pm 0.003$ & $0.264 \pm 0.002$ \\
		b & $1.301 \pm 0.043$ & $0.643 \pm 0.007$ & $0.222 \pm 0.007$ \\
		c & $1.441 \pm 0.103$ & $0.776 \pm 0.045$ & $0.283 \pm 0.015$ \\
		d & $4.352\pm 0.104$ & $1.556 \pm 0.019$ & $0.434 \pm 0.010$\\
		\hline	
		\multicolumn{4}{c}{distance constrained}\\
		b & $1.324 \pm 0.027$ & $0.645 \pm 0.006$ & $0.221 \pm 0.007$ \\
		d & $4.150 \pm 0.066$ & $1.543 \pm 0.020$ & $0.441 \pm 0.010$ 
	\end{tabular}
\end{table}

\begin{figure*}[htbp]
	\includegraphics[width=0.49\hsize]{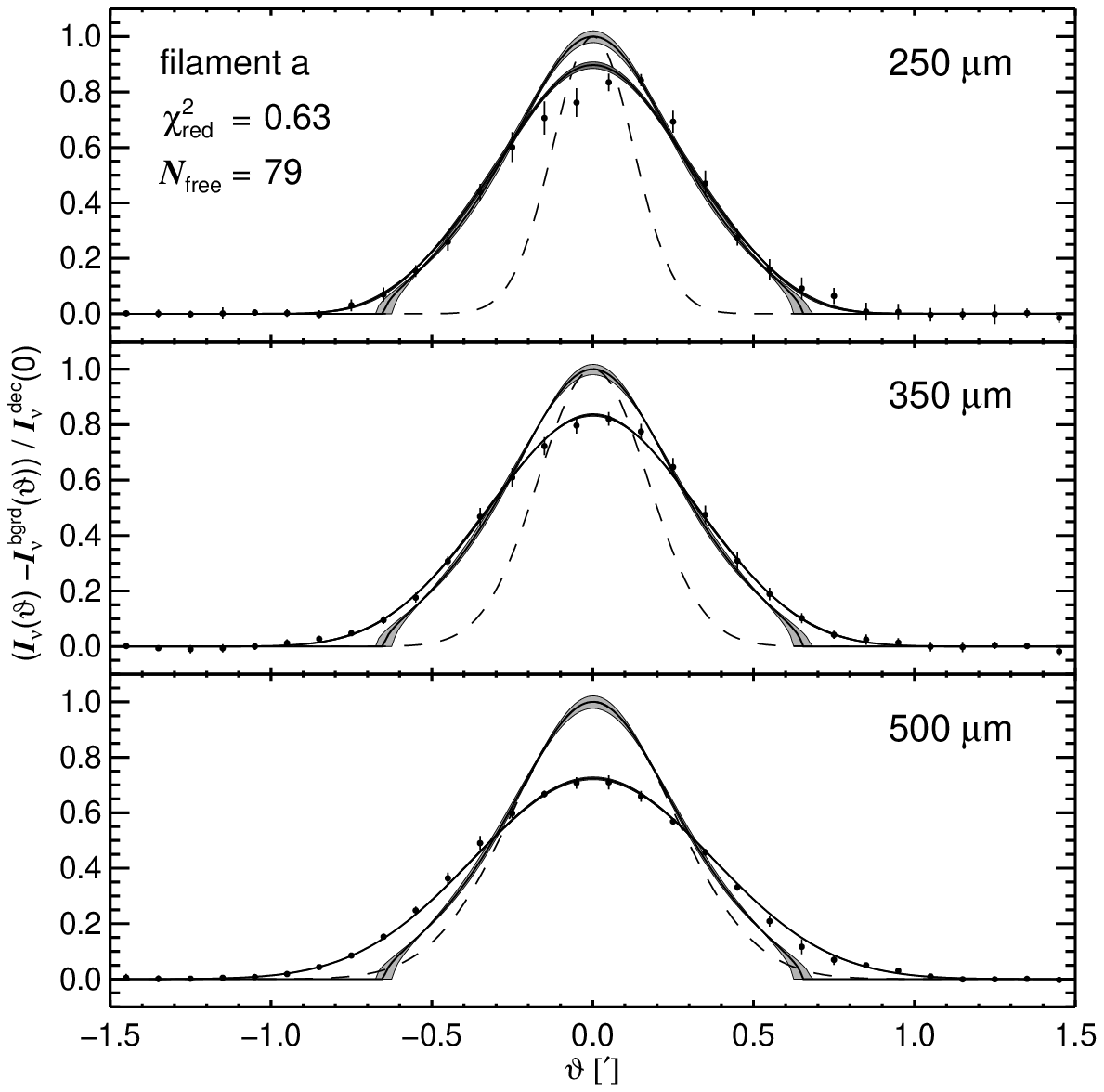}
	\hfill
	\includegraphics[width=0.49\hsize]{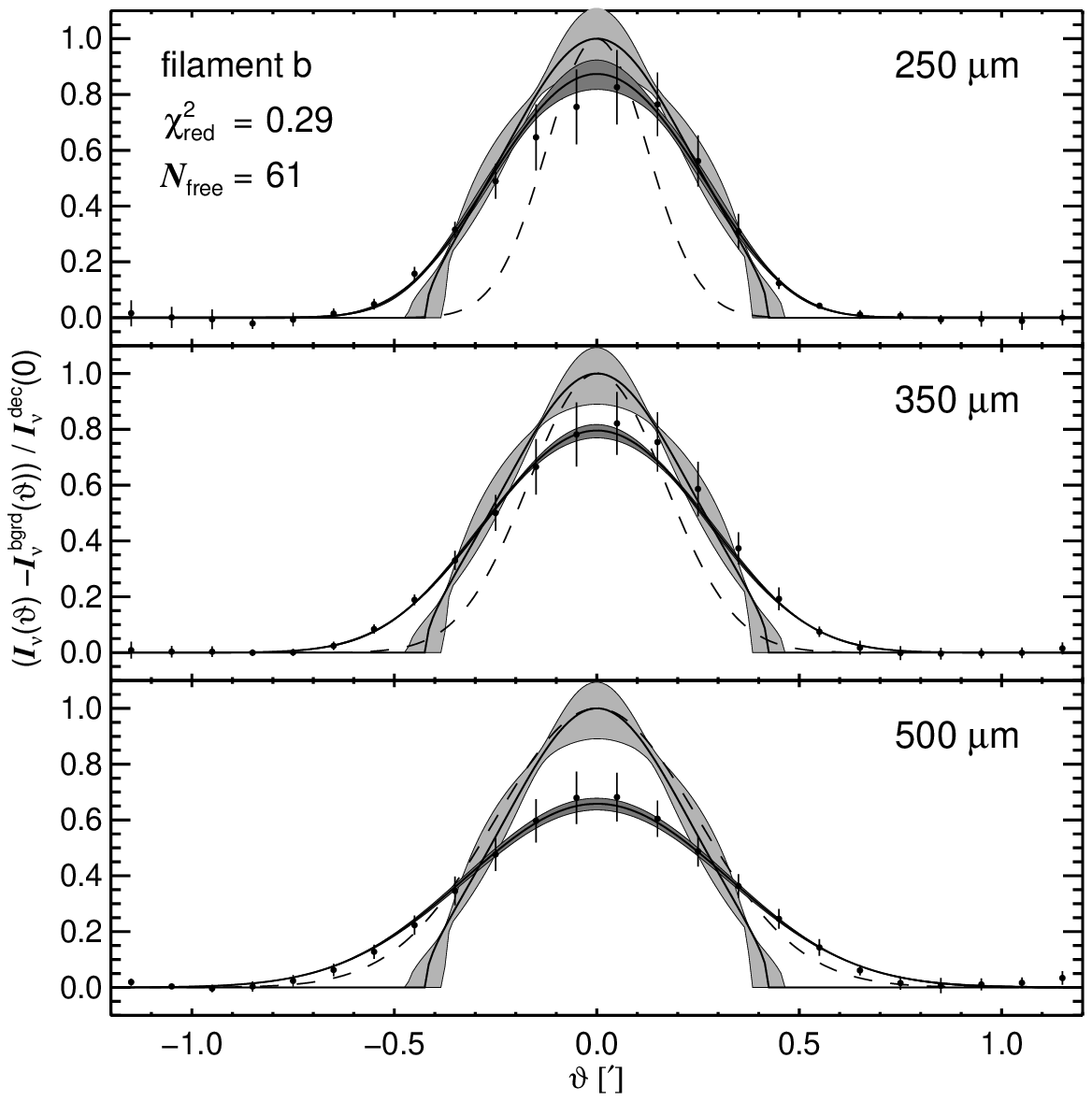}

	\includegraphics[width=0.49\hsize]{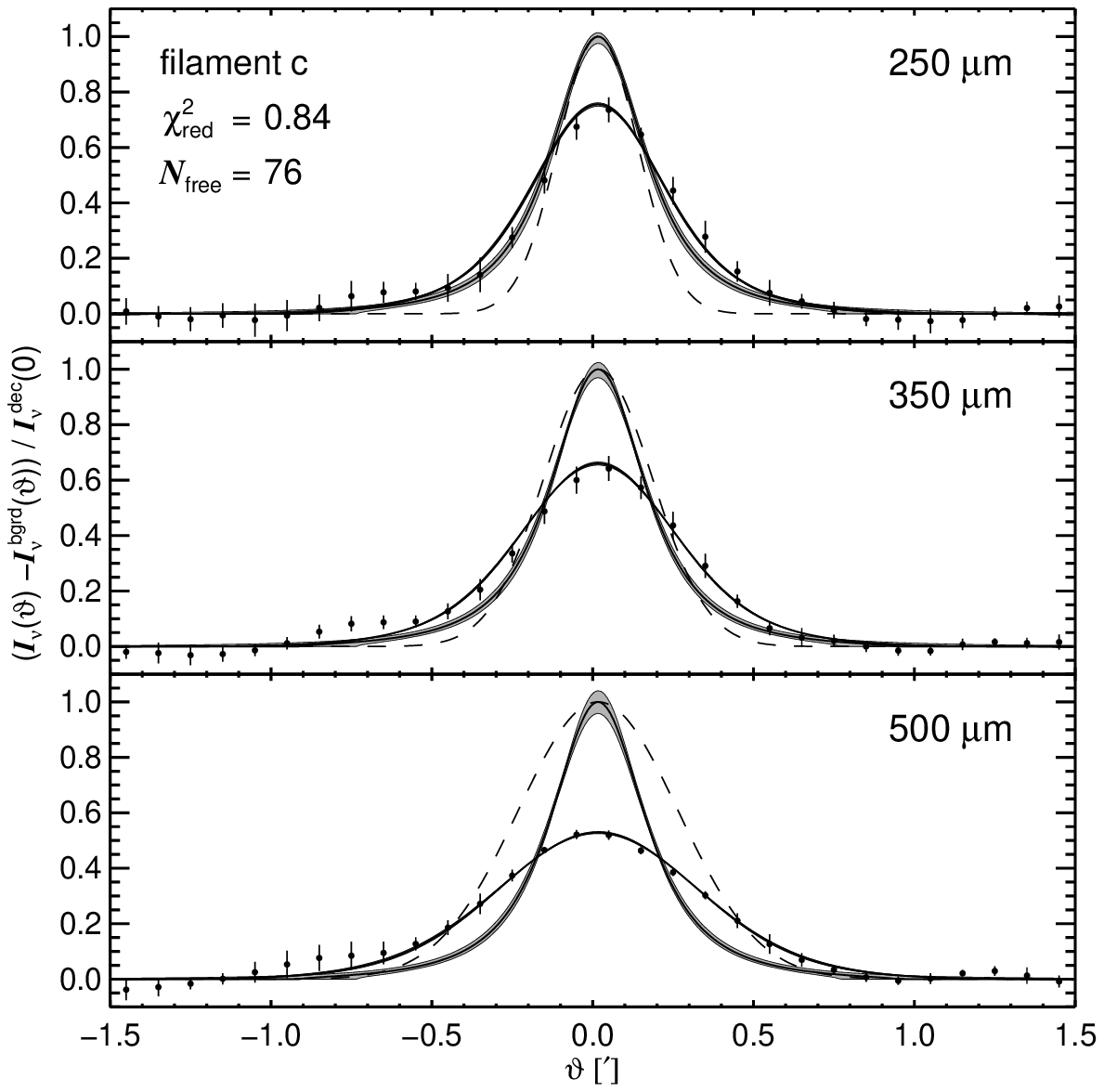}
	\hfill
	\includegraphics[width=0.49\hsize]{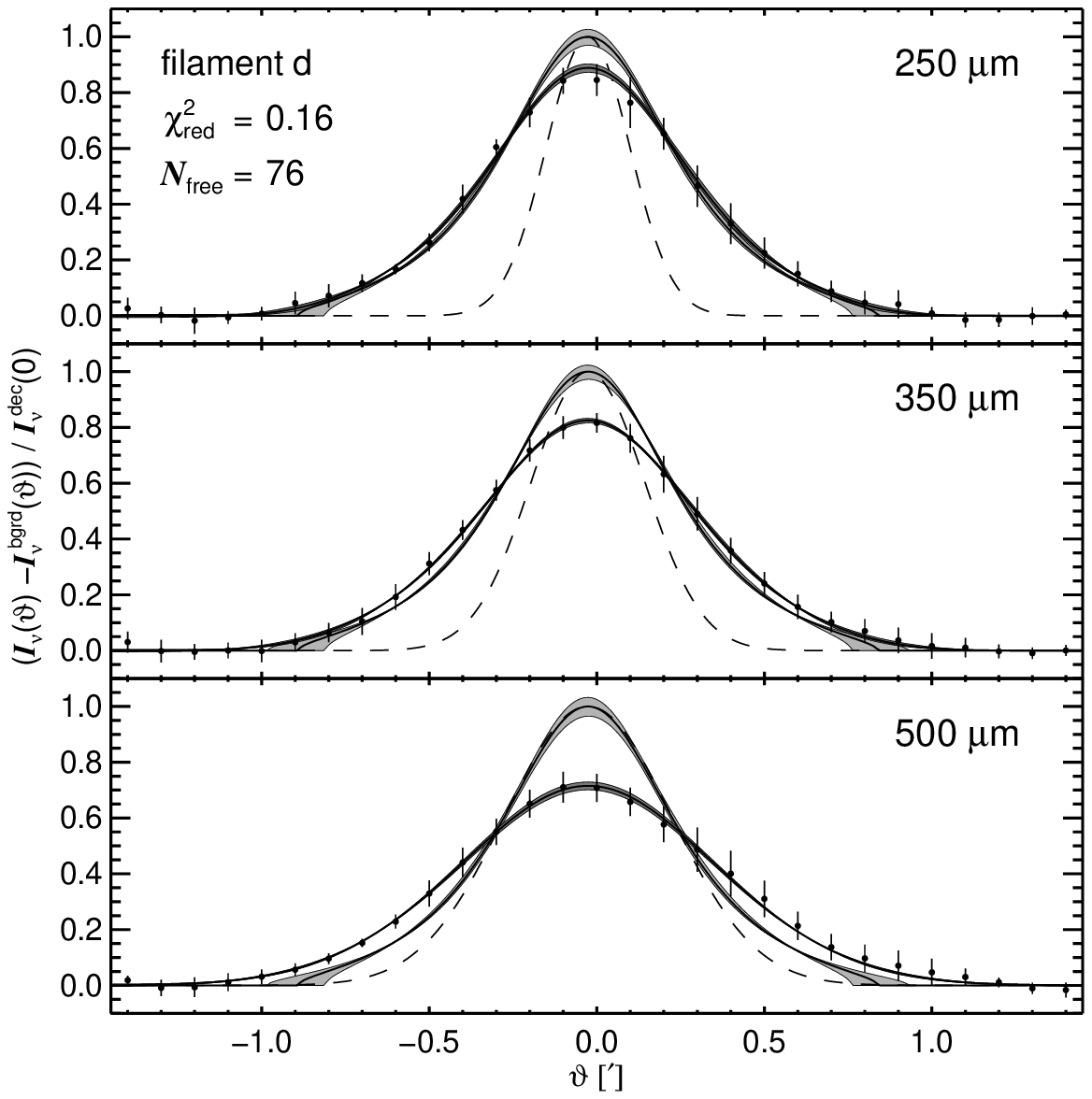}
	\caption{\label{fig_profile} 
	Surface brightness profiles of the filaments  at 250, 350, and $500~\mu{\rm m}$, normalized by the central surface 
	brightness of the intrinsic model, before convolution.  The model is optimized so that after convolution it matches the data.  
	Note that the multi wavelength data simultaneously fit the SED (Fig.~\ref{fig_sed}). 
	The light grey shaded area shows the 1 sigma variations of the intrinsic surface brightness profiles of the model.  
	The best fit model of the data has a peak less than unity because of convolution with the beam; the 1 sigma 
	variations (most easily seen for filament $b)$) are shown by the dark areas.  The data that were fit, corrected using the best fit for the background 
	emission, are shown as points with error bars normalized to the best fit model. The adopted Gaussian approximations 
	of the psf are given as dashed lines. North is to the right. Note that the scan lengths vary from filament to filament. 
	}
\end{figure*}

\begin{figure*}[htbp]
	\includegraphics[width=0.49\hsize]{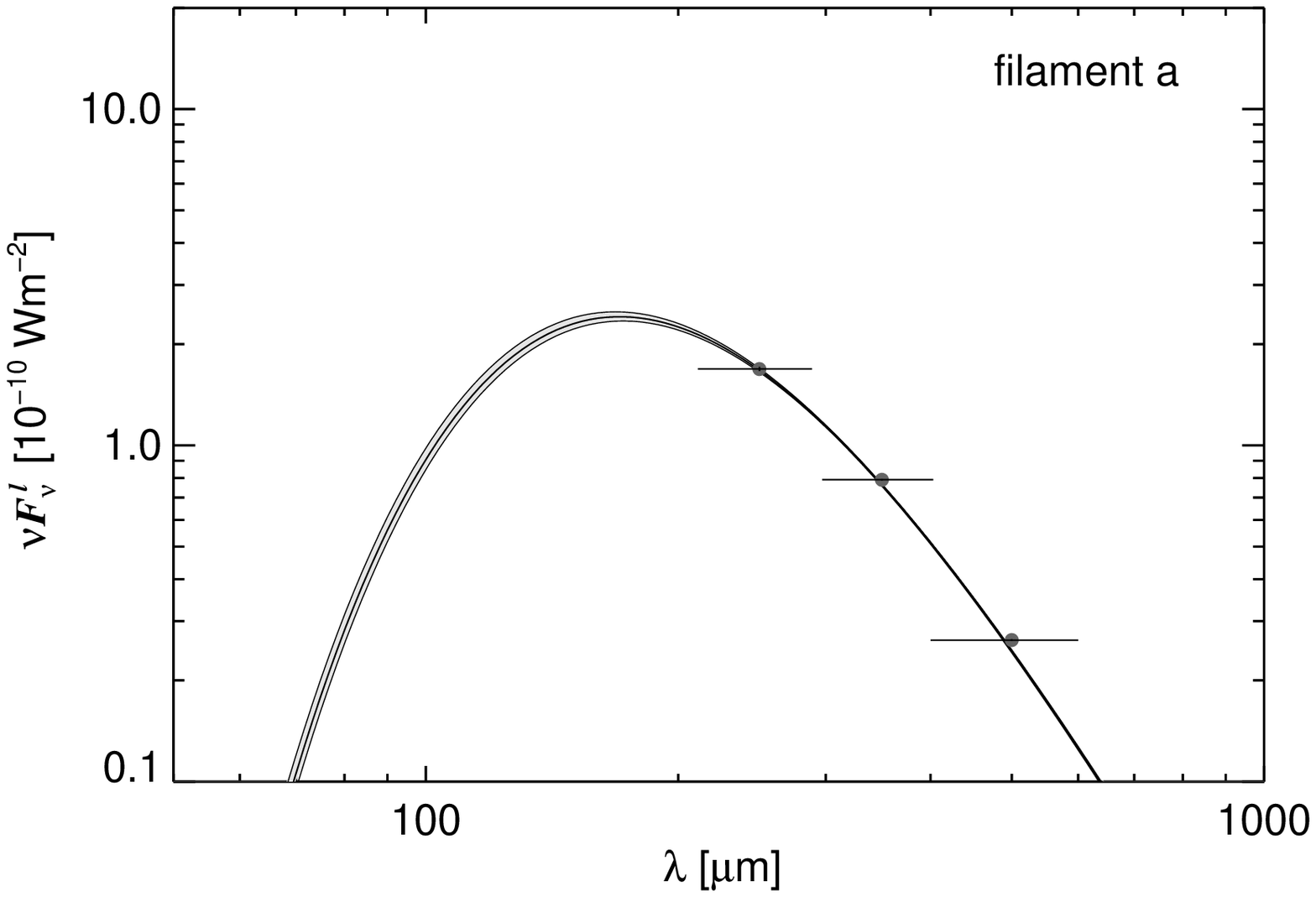}
	\hfill	
	\includegraphics[width=0.49\hsize]{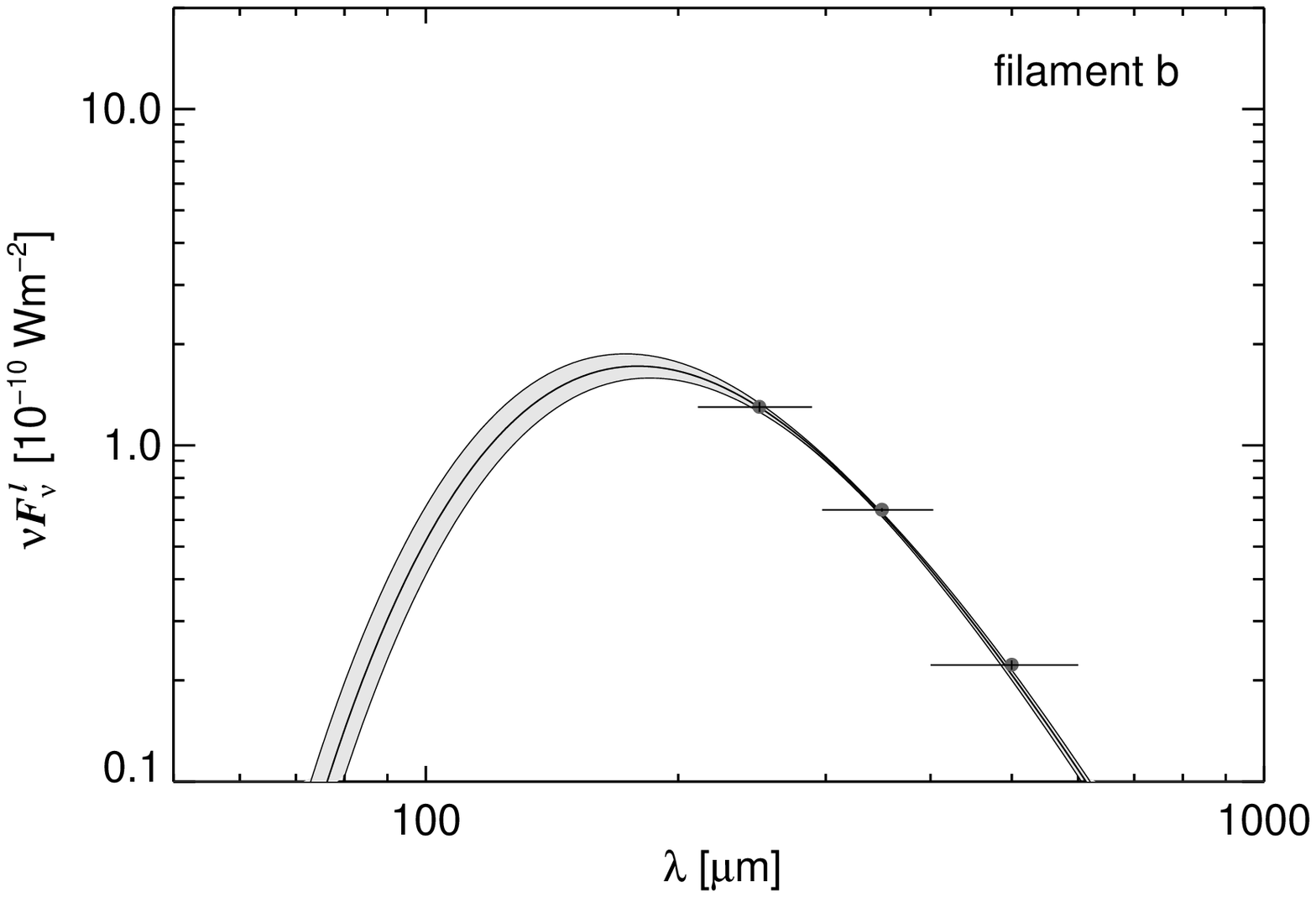}

	\includegraphics[width=0.49\hsize]{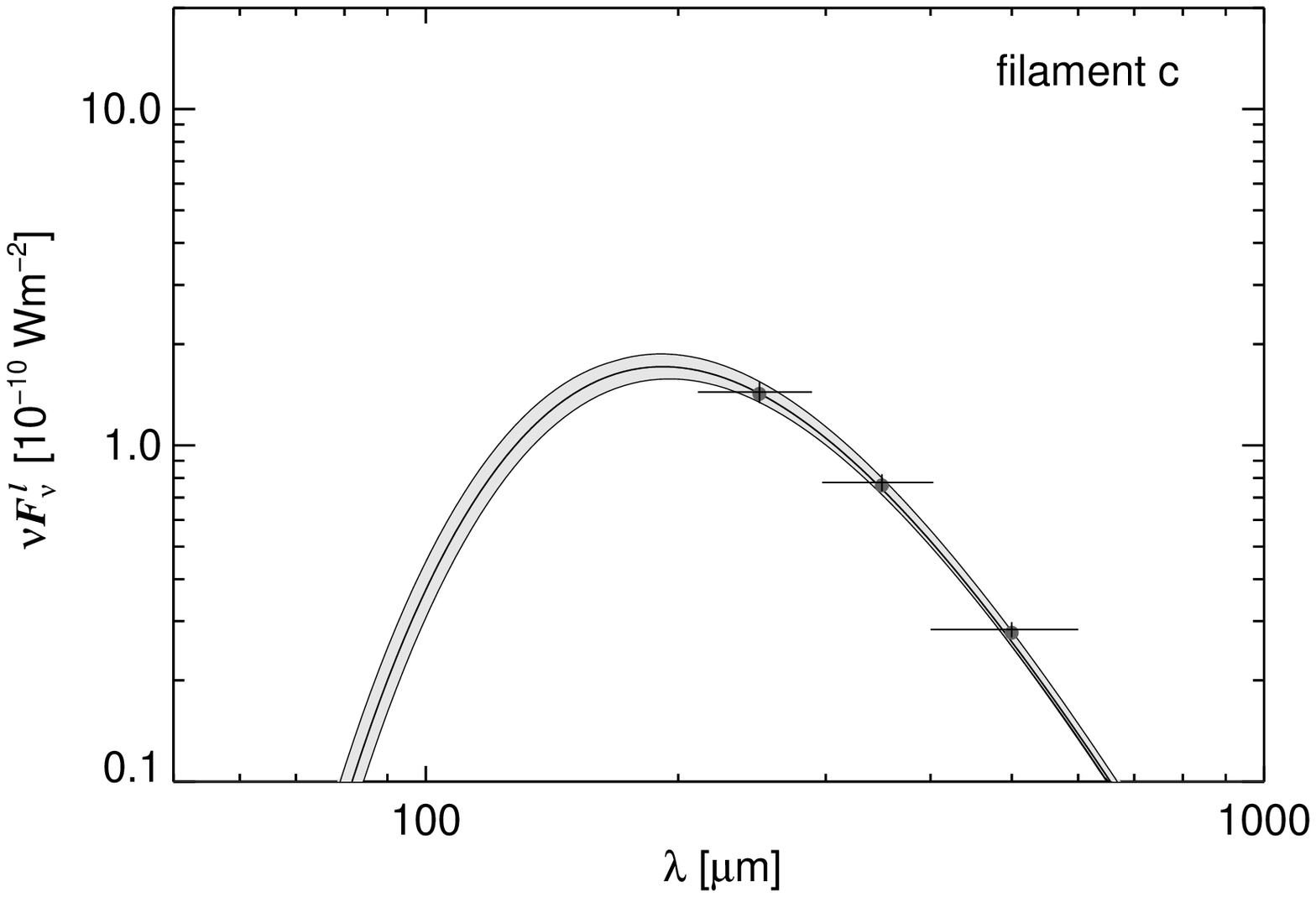}
	\hfill
	\includegraphics[width=0.49\hsize]{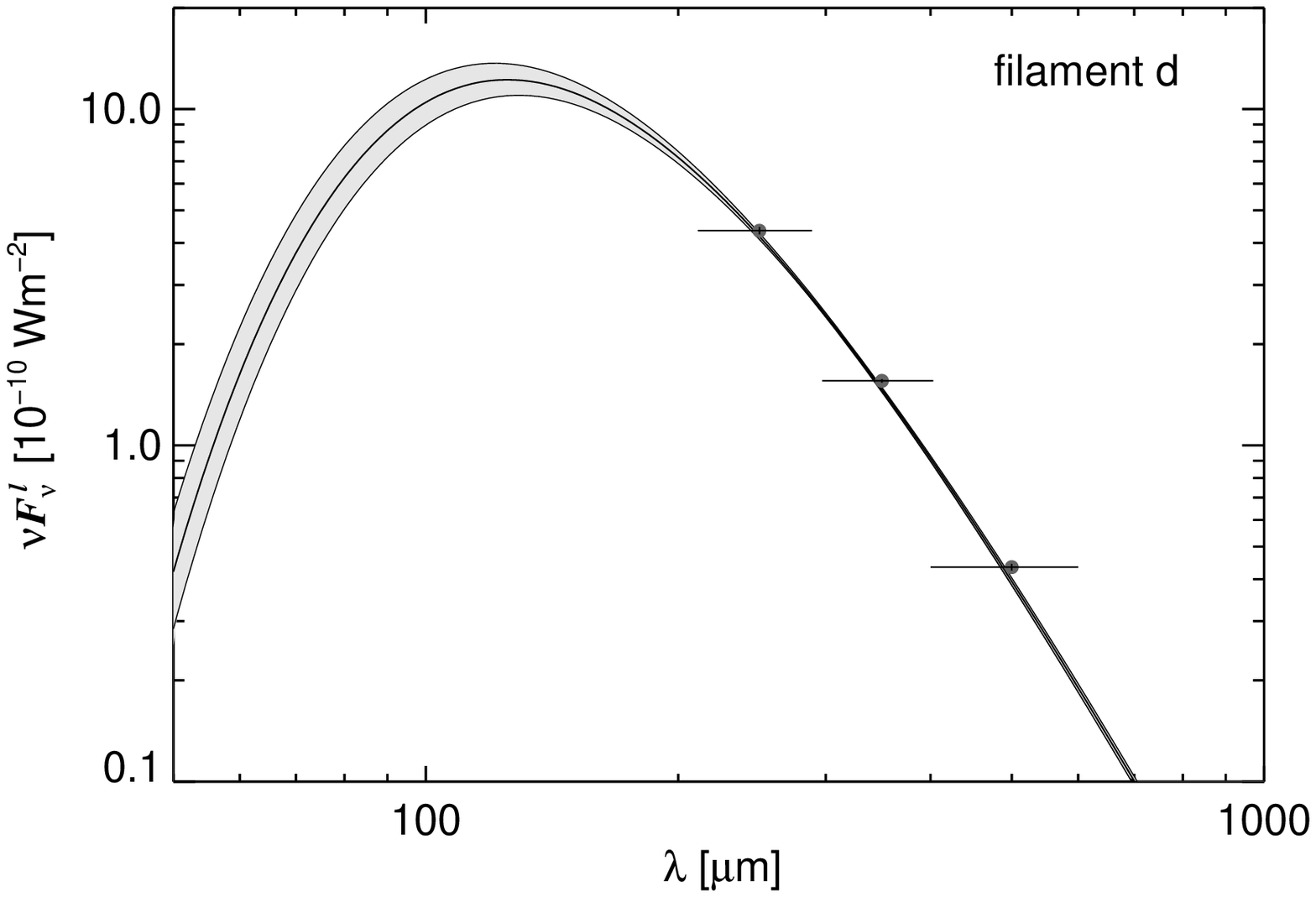}
	\caption{\label{fig_sed}
	Spectral energy distributions for the four filaments
	derived by a simultaneous fit of the profiles for unconstrained distances
	assuming for the dust emission spectra a modified black body.
	The theoretical flux densities of the SPIRE filters (grey filled circles) correspond
	to the best fit (black line). The vertical lines mark
	the uncertainties of the theoretical fluxes of the broad band filters and the horizontal lines the filter widths.  
	The grey shaded areas show the $1 \sigma$ uncertainties of the SED.
	}
\end{figure*}

Overall, we found a good agreement of the theoretical and the observed
surface brightness profiles with a reduced $\chi^2_{\rm red}<1$
(Fig.~\ref{fig_profile}) in all four cases.  The derived parameters
for the four filaments are listed in Tab.~\ref{table_fitmodel} and
Tab.~\ref{table_fitmodel2}.
The temperatures for the filaments are close to those derived by
\citet{Arzoumanian2011}.  They are larger than the adopted $T_{\rm
cyl}$.  As discussed in \citetalias{Fischera2012a}, efficient coupling is not expected in
these filaments, especially in the outskirts which dominate in
determining $T_{\rm dust}$.  Because $T_{\rm cyl}$ is a scaling
parameter of the model, the choice does not affect the quality of fit
of the model, but it systematically affects the parameters
(Sect.~\ref{sect_discussion}.

The fluxes of the broad band filters given in Tab.~\ref{table_fluxes}
are derived using Eq.~\ref{eq_flux} with the fitted results for $f_{\rm cyl}$, $D\cos i$, and $T_{\rm dust}$ (Fig.~\ref{fig_sed}).
The three photometric images appear to be sufficiently accurate for a first analysis of the 
physical parameters of the filaments despite the incomplete sampling of the SED.
The largest systematic uncertainty is certainly expected for filament d. But even in this case the SED seems to be well determined.

For comparison we also assumed a Gaussian profile which produced
similar flux estimates.  The derived parameters for the dust
temperature, size, and central column density were all similar to the
ones we obtained by a proper model of the emission profile.  Larger
deviations were found for filament c where the Gaussian approximation
leads to a systematically lower flux and an apparent larger size
(\FWHM{}).  With this Gaussian approximation, however, we gain no
information on $D$, or $p_{\rm ext}$.
In addition, in this approximation the mass ratio $f_{\rm cyl}$
can only be determined as the ratio of the derived mass line density
and the maximum mass line density $2 K/G$, the former uncertain by the
inclination angle and the distance $D$ and the latter uncertain by
$T_{\rm cyl}$.  In the model fit presented here $f_{\rm cyl}$ is determined
from the shape of the profile.

\begin{figure*}[htbp]
	\includegraphics[width=0.49\hsize]{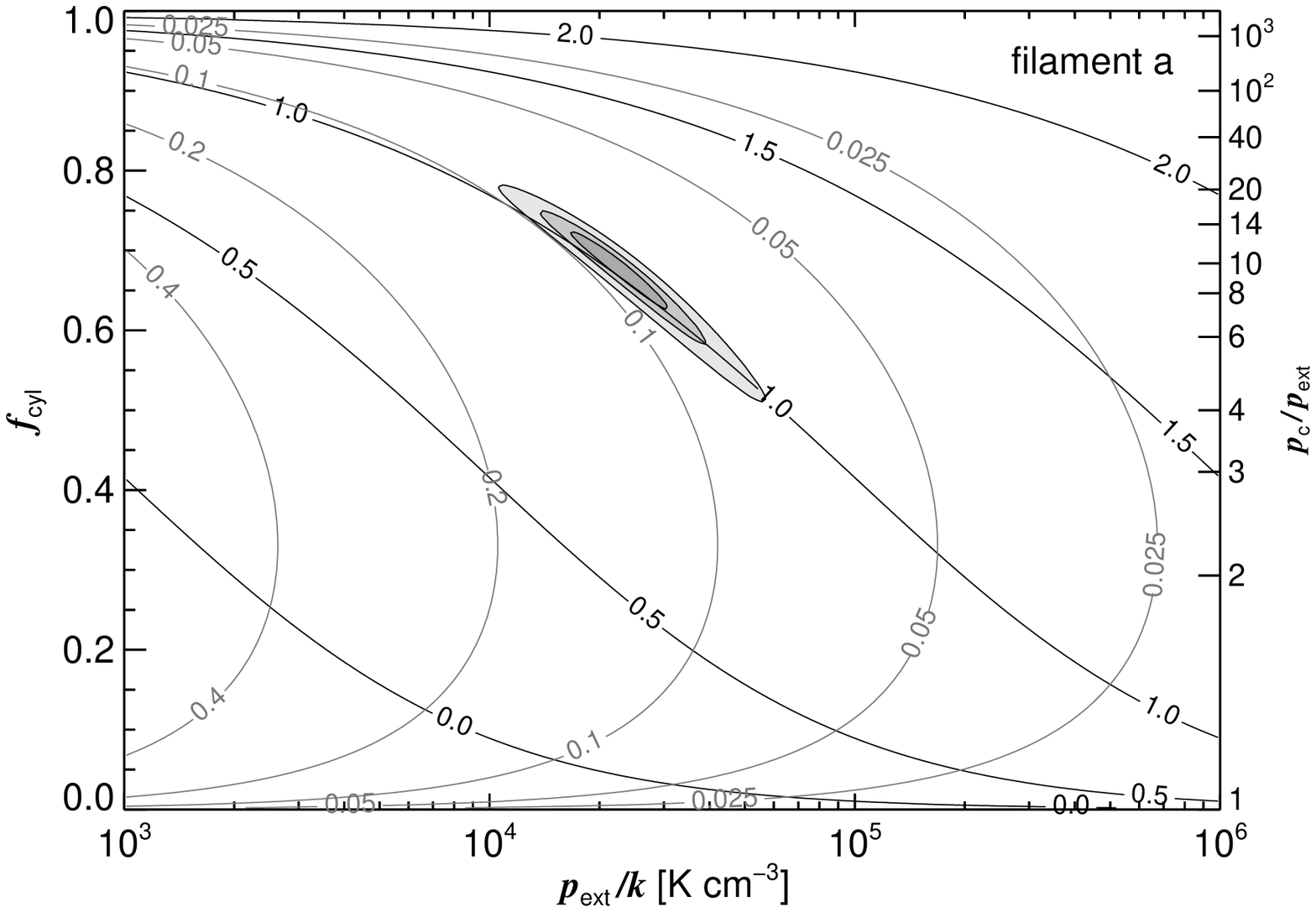}
	\hfill
	\includegraphics[width=0.49\hsize]{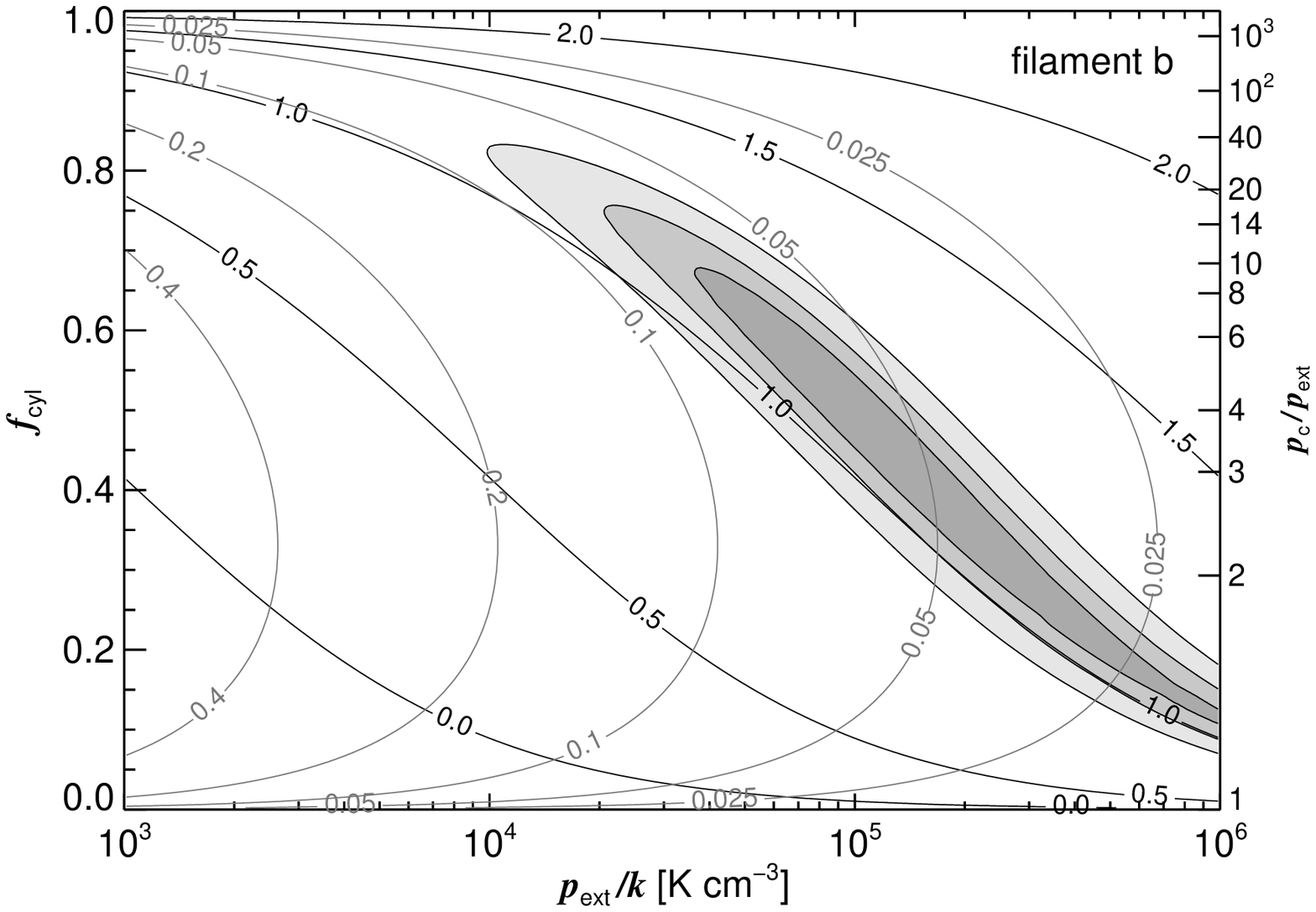}
	
	\includegraphics[width=0.49\hsize]{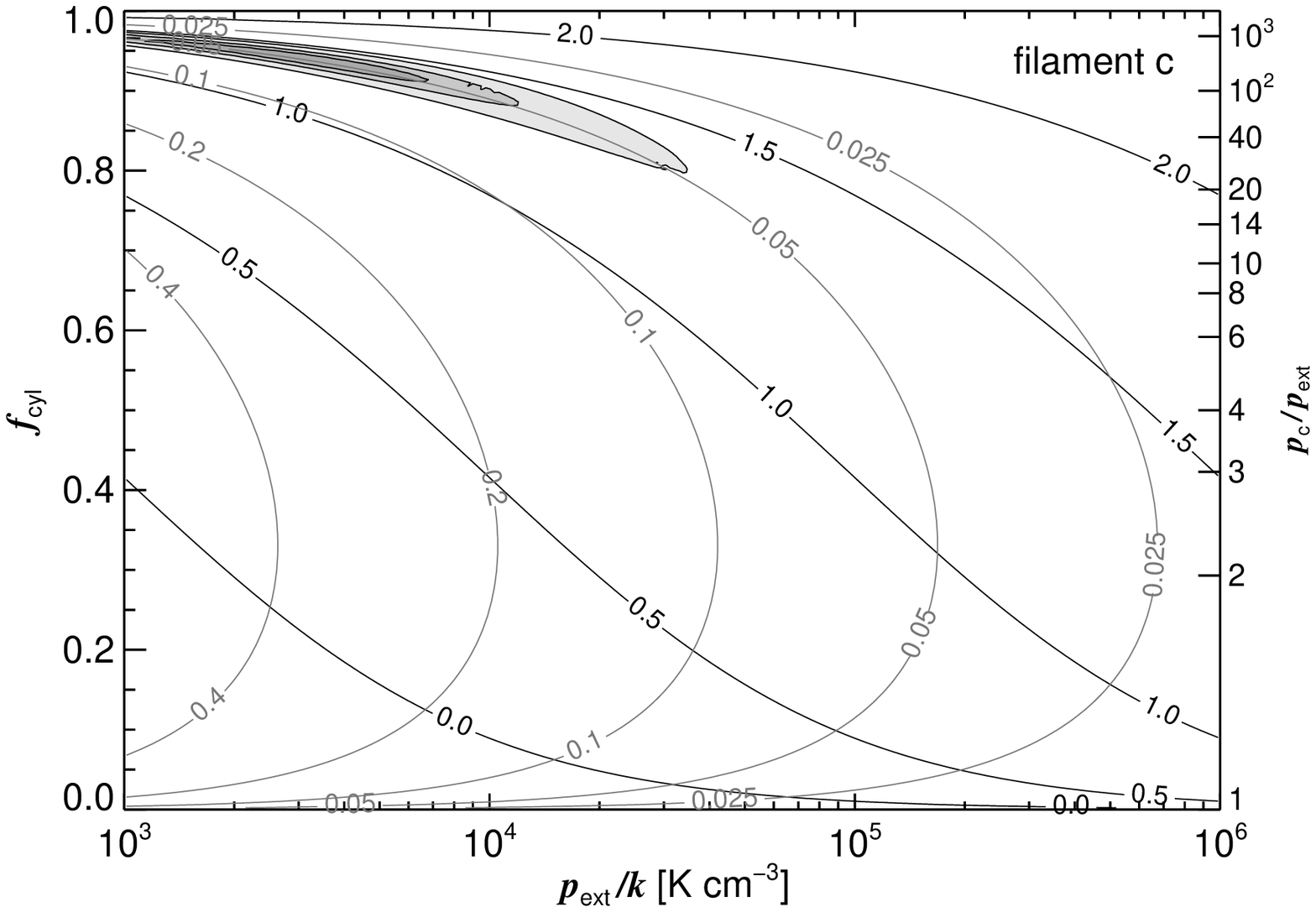}
	\hfill
	\includegraphics[width=0.49\hsize]{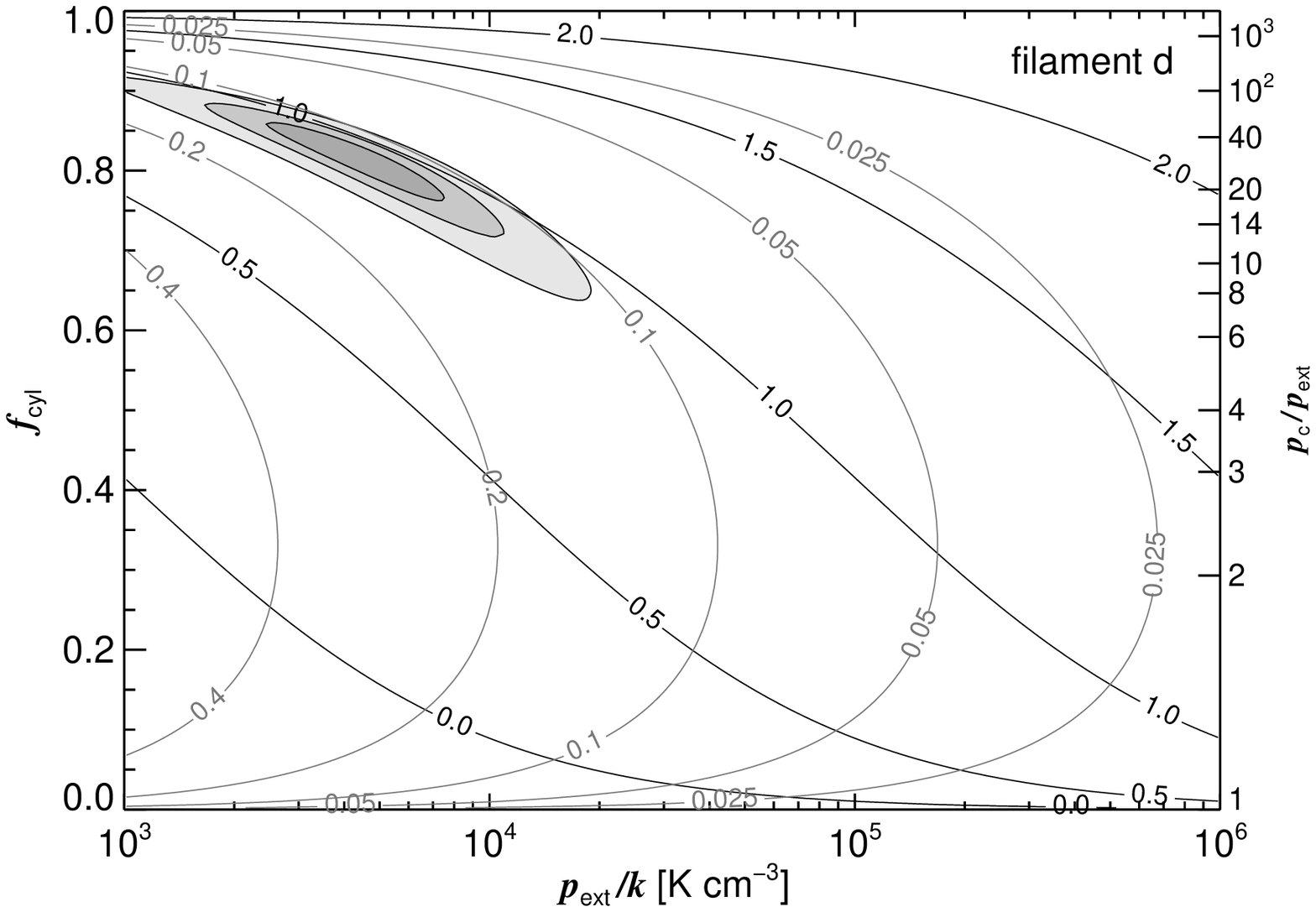}
	\caption{\label{fig_chisqr} Regions of highest confidence  in the $f_{\rm cyl}$-$p_{\rm ext}$ plane
	for the fit of the surface density profile at 250, 350,
	and $500~\mu{\rm m}$ with the model of an isothermal self-gravitating pressurized 
	infinite cylinder (dark grey: 68\%, medium grey: 90\%, light grey: 99\%). 
	The cylinder is assumed to be seen edge-on. 
	The corresponding overpressure $p_{\rm c}/p_{\rm ext}=(1-f_{\rm cyl})^{-2}$ is labeled on the right axis. 	
	The black curves give the column densities
	of neutral hydrogen through the centre, labelled with $\log_{10}N_{\rm H}(0)~[10^{21}{\rm cm}^{-2}]$. The
	grey lines give the \FWHM{} in parsec. 
	}
\end{figure*}

The result of the detailed model for the external pressure and the mass ratio $f_{\rm cyl}$ is visualized in
Fig.~\ref{fig_chisqr}. We note that the regions of highest probability follow lines of constant central column
density. 
The lines of constant central column densities are given by
\begin{equation}
	N_{\rm H}(0) \approx \frac{1}{\bar \mu m_{\rm H}}\sqrt{\frac{p_{\rm ext}}{4\pi G}}2\sqrt{8}\sqrt{f_{\rm cyl}},
\end{equation}
at low mass ratios and by
\begin{equation}
	\label{eq_columnhighf}
	N_{\rm H}(0) \approx \frac{1}{\bar \mu m_{\rm H}}\sqrt{\frac{p_{\rm ext}}{4\pi G}}
			\frac{\sqrt{8}}{1-f_{\rm cyl}}\frac{\pi}{2},
\end{equation}
at high mass ratios $f_{\rm cyl} >\approx 0.7$. 

In the figure we added lines of constant size (\FWHM{}).
For given external pressure and cloud temperature the size varies as a function of mass ratio $f_{\rm cyl}$.
At both low and high mass ratios the size approaches zero, at low mass ratios because of the vanishing 
mass and at large mass ratios because of compression. At $f_{\rm cyl}\approx 0.331$ a cylinder has the
maximum size at a given $p_{\rm ext}$. The lines of constant \FWHM{} are given through the equations (\citetalias{Fischera2012a})
\begin{equation}
	\FWHMEQ  \approx 
		 \sqrt{3}\frac{\sqrt{8}K}{\sqrt{4\pi G p_{\rm ext}}}\sqrt{f_{\rm cyl}},
\end{equation}
for $f_{\rm cyl} < \approx 0.1$		 
and 
\begin{equation}
	\label{eq_fwhmhighf}
	\FWHMEQ \approx	2\sqrt{2^{2/3}-1}\frac{\sqrt{8}K}{\sqrt{4\pi G p_{\rm ext}}} (1-f_{\rm cyl}),
\end{equation}
for $f_{\rm cyl} > \approx 0.7$.

At high mass ratios for both constant \FWHM{} and $N_{\rm H}(0)$ the functional dependence
of $p_{\rm ext}$ on $f_{\rm cyl}$ is identical ($p_{\rm ext} \propto (1-f_{\rm cyl})^2$)
and so the corresponding loci become parallel in Fig.~\ref{fig_chisqr}.

\begin{figure*}[htbp]
	\includegraphics[width=0.49\hsize]{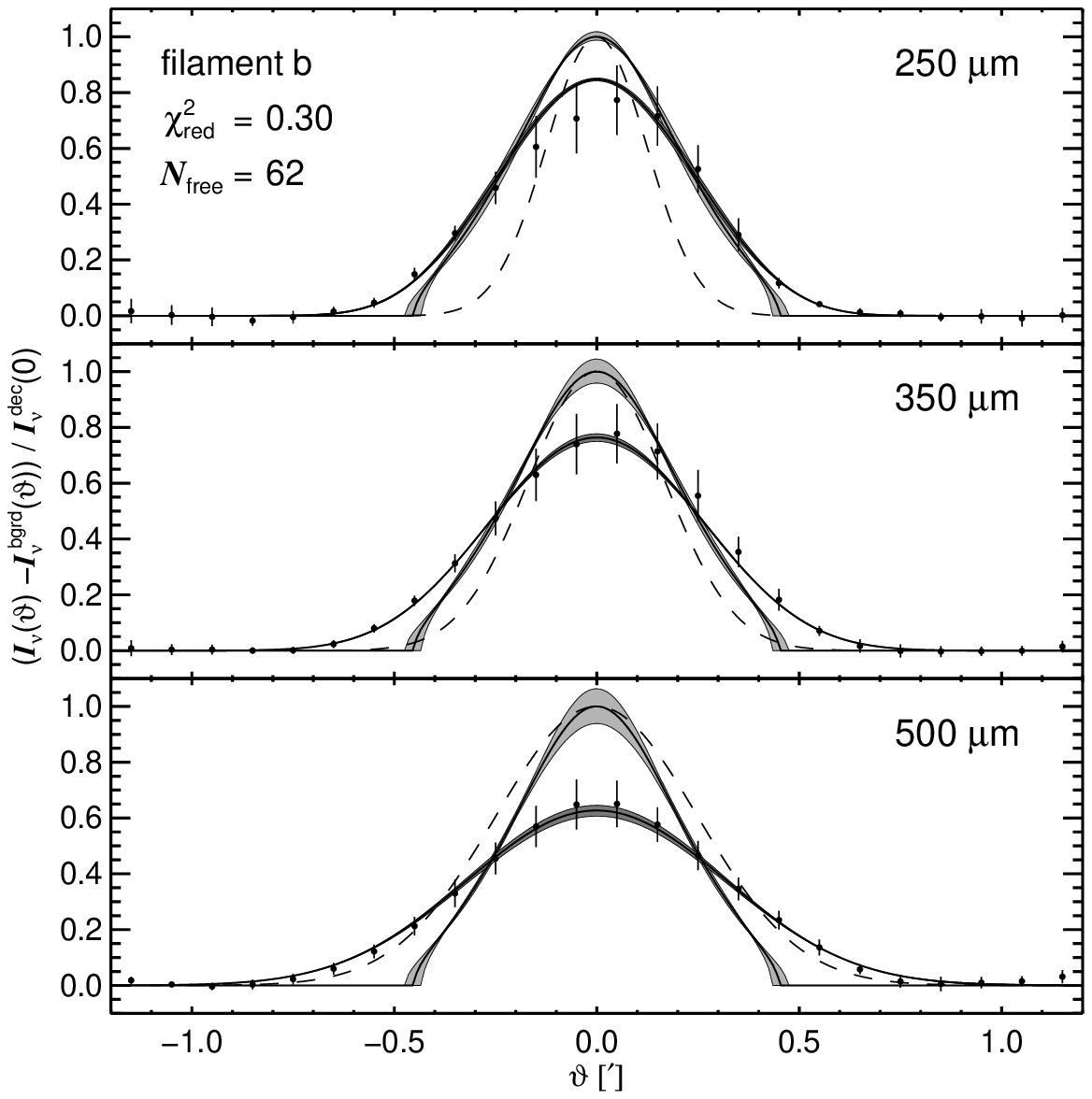}
	\hfill
	\includegraphics[width=0.49\hsize]{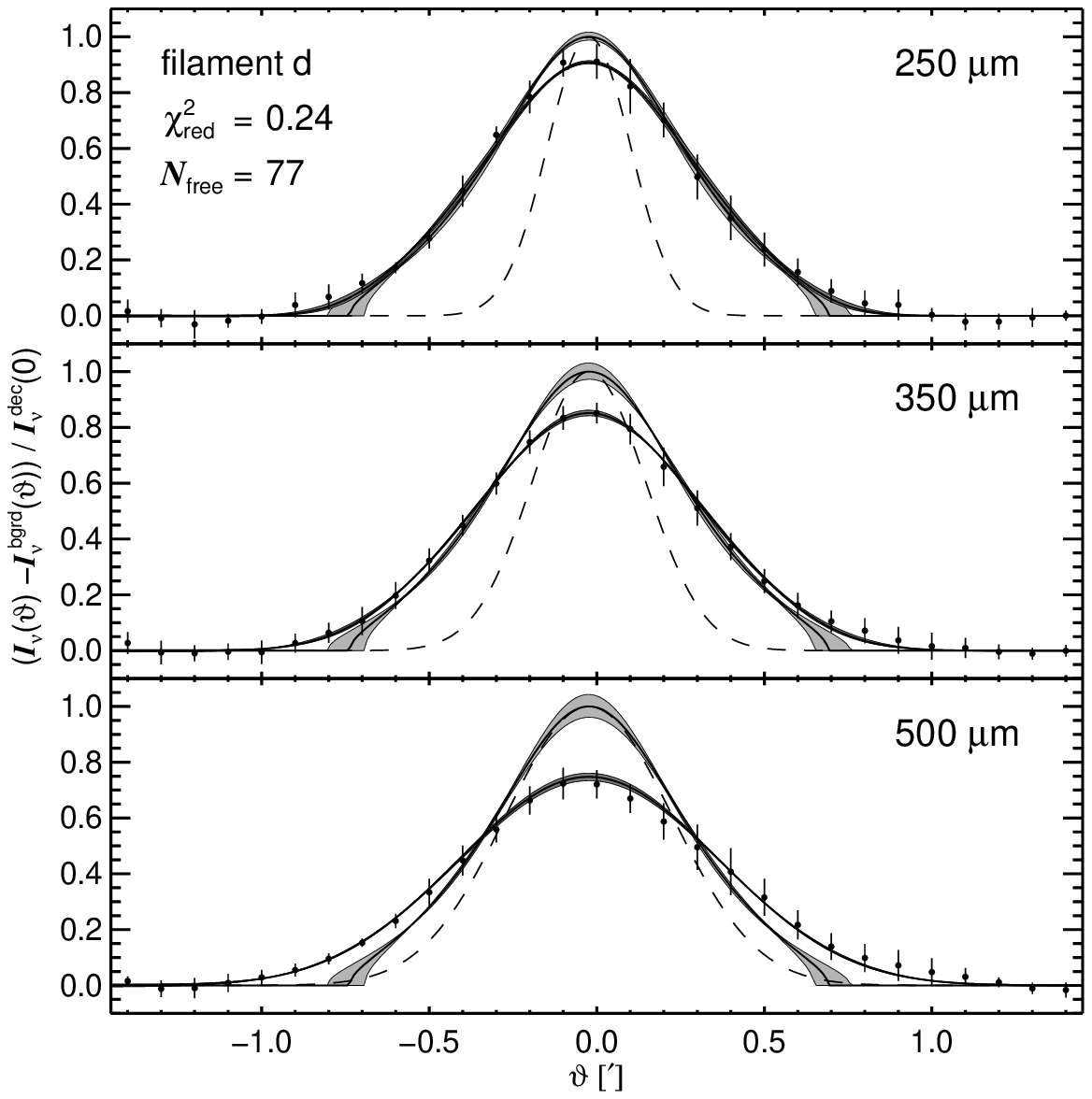}
	\caption{\label{fig_profile2} Same as Fig.~\ref{fig_profile} but  for constrained distance $493\pm 36$~pc.}
	\includegraphics[width=0.49\hsize]{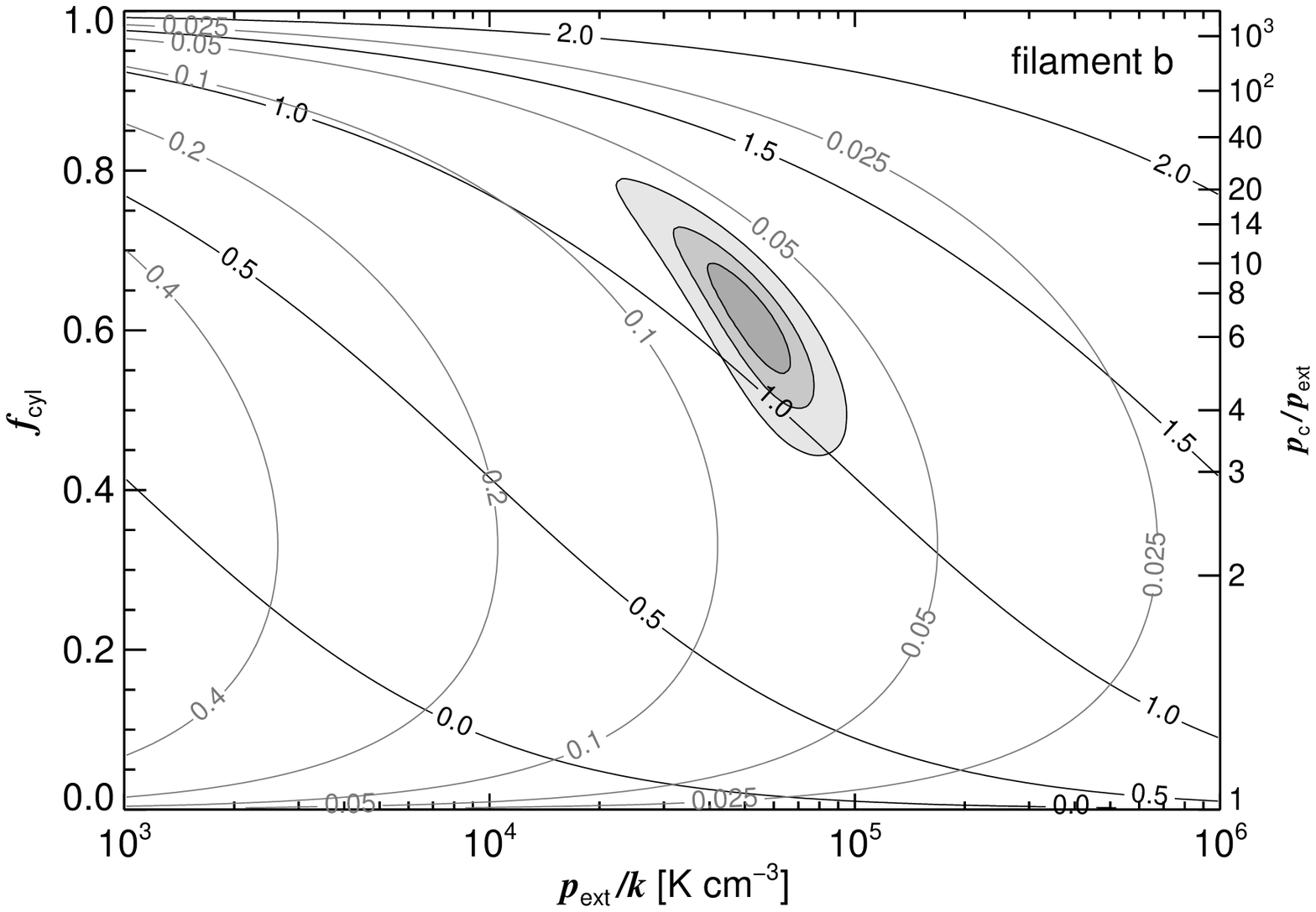}
	\hfill
	\includegraphics[width=0.49\hsize]{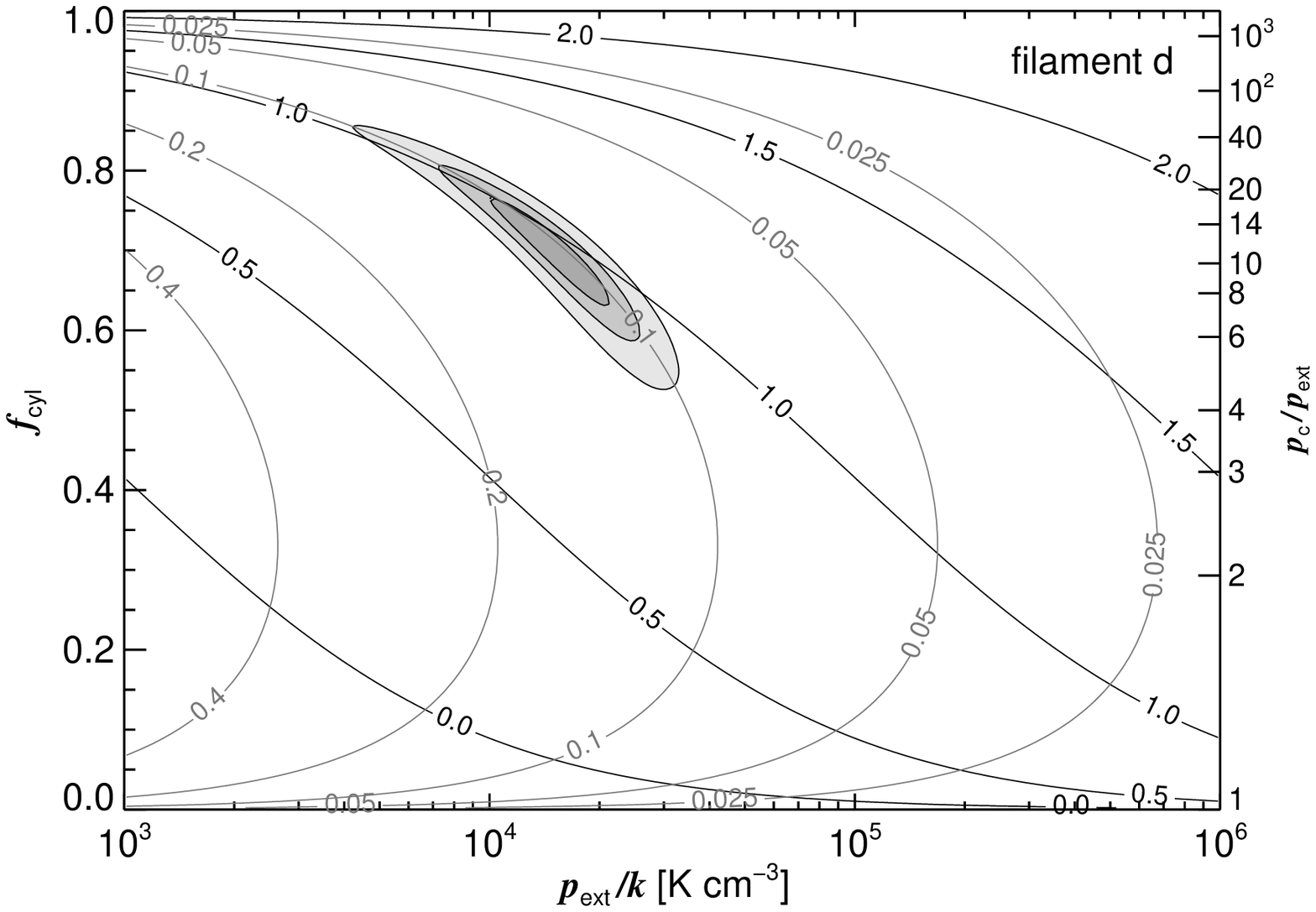}	
	\caption{\label{fig_chisqr2} Same as Fig.~\ref{fig_chisqr} but for constrained distance $493\pm 36$~pc.}
\end{figure*}

The estimate of all three parameters (distance $D$, external pressure $p_{\rm ext}$, and mass ratio $f_{\rm cyl}$)
requires accurate observations of the surface brightness profiles. 
To lower the uncertainty it is reasonable to use an additional constraint that the cloud distance should be roughly 
the same for all filaments. For filaments b and d we performed additional model fits 
where we constrained the distance using the estimate derived for filament a; the distance is not fixed but 
allowed to vary within the $1\sigma$ range. The corresponding fitted parameters are listed in 
Tabs.~\ref{table_fitmodel} and \ref{table_fitmodel2} and the derived fluxes in Tab.~\ref{table_fluxes}. 
As shown in Fig.~\ref{fig_profile2} and Fig.~\ref{fig_chisqr2} the additional constraint leads 
to tighter confidence intervals for the pressure and the mass ratio.

The filaments have deconvolved sizes (\FWHM{}) 
on the order of $0.1~\rm pc$. 
All column densities are relatively high ranging from
$N_{\rm H}(0)\sim 8\times 10^{21}~{\rm cm^{-2}}$ to $\sim 23\times 10^{21}~{\rm cm^{-2}}$. 
Filaments with higher central column density $N_{\rm H}(0)$ tend to have both a smaller \FWHM{} 
and steeper density profile, characterized though a higher $f_{\rm cyl}$. This behavior is consistent
with what is expected from the model of self-gravitating isothermal pressurized cylinders with mass ratios
$f_{\rm cyl}>0.331$. The $N_{\rm H}(0)$-\FWHM{} relation is discussed in Sect.~\ref{sect_systematics}.

Comparing filaments a, b, and c we find a decrease of dust temperature with column density as expected for
opaque filaments heated by the ambient radiation field. Strong radiative transfer effects are also indicated
through the $L/M$ ratio with values considerably lower than $1.15~L_\odot/M_{\odot}$, the typical ratio for diffuse
dust heated by the interstellar radiation field \citep{Fischera2011}. 
The filaments a, b, and c have cold dust temperatures
below $15~{\rm K}$ in agreement with what is expected for dust grains in the molecular phase.
The warmer dust temperature and the higher $L/M$ ratio for filament d indicate strong heating from recently formed stars in the 
Cocoon Nebula. The radiation field is probably therefore highly non isotropic.

The four filaments indicate a trend in which more opaque filaments are also more compact as expected
for filaments with strong self-gravity where $\FWHMEQ\propto 1/N_{\rm H}(0)$ (Fig.~\ref{fig_resultgauss}). 

In the following we will describe the results for the four filaments in more detail.

\subsection{Filament a}

This filament provides the best estimate of the physical parameters because of the well defined background
and source emission as seen in Fig.~\ref{fig_profile}. The surface brightness profile is well reproduced with the model. The only apparent 
deviation can be seen in the interior of the profile at $250~\mu{\rm m}$ which is asymmetric.
This might be related to substructure below the resolution.
Another possibility is the temperature variation caused by radiative effects and a non isotropic radiation field
which is expected if the cloud complex as pointed out in Sect.~\ref{sect_observations} is located below the 
Galactic plane.
In particular the UV and optical radiation
could therefore be stronger at the surface facing the Galactic plane (right side in Fig.~\ref{fig_profile}) 
leading to warmer dust temperatures.
A similar asymmetry is seen in the surface brightness profiles of filament c.
The origin of these asymmetries might be resolved using PACS data.

\subsection{Filament b}

The filament b shows the strongest variation of the surface brightness along the filament axis.
A closer examination suggests that the variations are related to a condensation
in the filament centre. 
Applying the method described in App.~\ref{app_source} for the considered 
scan aperture we extracted the source emission.
The source is elongated along the filament with axial ratio 2 and a length of $19.3''\pm 0.7''$.
The dust temperature with $T_{\rm dust}=12.7\pm 0.3$ is considerably colder than obtained 
for the filament. The mass is $1.01\pm 0.07~M_\odot$. 


Because of the larger uncertainties a precise
determination of all the parameters, distance, pressure, and mass ratio, is not possible. 
Fig.~\ref{fig_chisqr} shows therefore a very extended confidence region. The distribution, however,
is not random but follows a line of constant central column density. 

The fitted distance varies systematically with $f_{\rm cyl}$ and $p_{\rm ext}$. For
given mass ratio $f_{\rm cyl}$ the distance decreases for higher pressure because the filaments are
intrinsically smaller in higher pressure regions (Eq.~\ref{eq_radius}). At low mass ratios ($f_{\rm cyl}<0.5$) the
radius decreases for given pressure with $f_{\rm cyl}$ which leads to a systematic 
decrease of the distance with decreasing $f_{\rm cyl}$.
Therefore, when we use the larger distance derived for filament a as a constraint we obtain more tightly
constrained intrinsic profiles (Fig.~\ref{fig_profile2}) and more tightly 
defined estimates of the gravitational state $f_{\rm cyl}$ and the external pressure (Fig.~\ref{fig_chisqr2}), near the higher
$f_{\rm cyl}$ and lower $p_{\rm ext}$ end of the original confidence range.

\subsection{Filament c}

Filament c is exceptional in several regards. It has the highest central column density,
the highest mass line density, the coldest dust temperature, and the smallest \FWHM{}.

The surface brightness profile indicates a filament with an extremely high mass ratio
with $f_{\rm cyl}>0.8$. The filament is therefore characterized by a high overpressure
$> 25$ and by an intrinsic density profile at the outskirts steeper than $r^{-2}$. As we can see
from Fig.~\ref{fig_profile}, at large impact parameters 
the profile differs quite substantially from a simple Gaussian profile, showing wide tails.

Even at the shortest wavelength the filament is not well resolved, with a size almost
identical to the angular resolution. Our model of the profile might therefore be affected to a 
larger degree by deviations of the true $psf$ from the Gaussian function used here for sake of simplicity.

An additional uncertainty is related to the source in the aperture that we subtracted
using the method described in App.~\ref{app_source}. The excess seen at $\vartheta\approx -0.75'$ 
in the background might be a residual of this source subtraction or a true structure. 
The interior profile shows a slight asymmetry with higher emission to the right. As in the case of filament a
that side faces towards the Galactic plane and the asymmetry might therefore be related
to radiative transfer effects. 


\subsection{\label{sect_filament_d}Filament d}

For this filament we obtain the largest distance. The value of $p_{\rm ext}$ seems to be more in agreement
with what is found for filament c. However, if we constrain the distance using the value derived for filament a 
we obtain for $p_{\rm ext}$ a value which is closer to that of filament a. The same applies for the mass ratio.


\section{\label{sect_discussion}Discussion}

\begin{figure}[htbp]
	\includegraphics[width=\hsize]{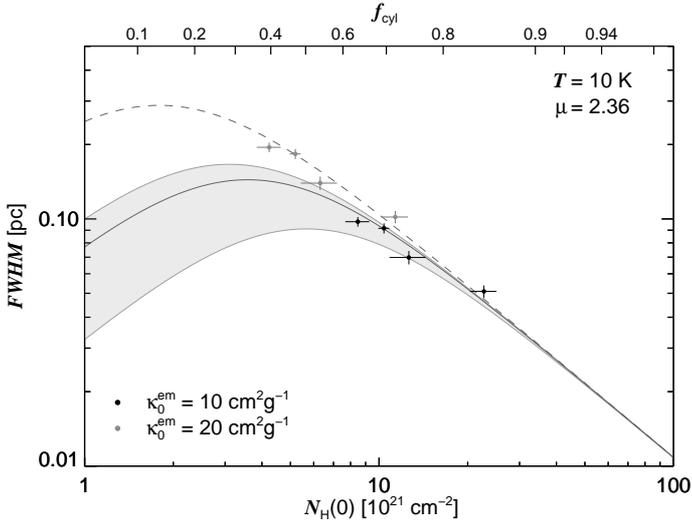}
	\caption{\label{fig_resultgauss}
	Relation between the \FWHM{}  and the central column density  (Tab.~\ref{table_fitmodel2}, col. 4 and 6)
	for the four filaments in \object{IC 5146}
	if the filaments are placed at the same distance $D=500~{\rm pc}$ (black circles) closely consistent
	with the assumed effective emission coefficient $\kappa_0^{\rm em}=10~{\rm cm^2 g^{-1}}$ and 
	an assumed dust-to-gas ratio $\delta=0.00588$. 
	For filaments b and d the values correspond to the fit with constrained distance.
	The curves of the relation shown are derived using the approximation (Eq.~32) provided in \citetalias{Fischera2012a}. 
	The grey shaded area shows the variation of the relation for a range of external pressures implied by 
	the observed sizes and column densities of filaments in Polaris, IC~5146, and Aquila \citepalias{Fischera2012a}.
	The external pressure is varied from $1.5\times 10^4$ to $5\times 10^4~{\rm K~cm^{-3}}$ where
	for given central column density the \FWHM{} is larger for lower pressure.
	The given mass ratios on the upper axis correspond to an external pressure $2\times 10^4~{\rm K~cm^{-3}}$ shown as 
	the solid black curve. Also shown are the observed properties if the effective emission coefficient is increased by
	a factor two (grey circles). The dashed curve shows the model relation for 
	$p_{\rm ext}/k=5\times 10^3~{\rm K~cm^{-3}}$.	
	}
\end{figure}

As we have shown the dust emission from filaments in combination with a physical model can provide 
valuable information about the filament itself. The profile or the gravitational condition (overpressure)
can be used to determine the pressure in the ambient medium and the distance to the filament. 

\subsection{\label{sect_systematics}Systematics}

The derived parameters for the distance and the pressure depend on the
assumed inclination angle and the dust emission coefficient. The
distance further depends on the effective temperature $T_{\rm cyl}$
which should 
ideally be determined in a separate observation using diagnostic
molecular line observations tracing gas of medium and high densities as used
in studies of condensed cores \citep{Rathborne2008}.
The model temperature $T_{\rm cyl}$ could be higher than the kinetic
temperature if the filaments are also supported by magnetic pressure
\citep{FiegePudritz2000a,FiegePudritz2000b}.  However, in \citetalias{Fischera2012a} we
did not find any strong indication for an elevated $T_{\rm cyl}$ that
would support this possibility. 
We note, that in general, although neglected in this paper, the fit
depends in addition on the exponent $\beta^{\rm em}$.

A lower emission coefficient is, according to Eq.~\ref{eq_emissivity} and Eq.~\ref{eq_surfbrightness},
compensated by a higher mass surface
density and therefore, because $\Sigma_{\rm M}\propto \sqrt{p_{\rm ext}}$ (Eq.~\ref{eq_profile}), 
by a higher pressure. A similar
effect is caused if the filament is seen at a projection angle $i>0$ (where $i=0$ refers to edge-on view).
Thus, the derived pressure varies systematically as $p_{\rm ext}\propto (\cos i / \delta\kappa^{\rm em}_0)^2$.

This systematic dependence has a further implication for our distance estimate.
According to Eq.~\ref{eq_radius} for a given mass ratio $f_{\rm cyl}$
the radius varies as $r_{\rm cyl}\propto T_{\rm cyl} / \sqrt{p_{\rm ext}}$. Since $D = s\times r_{\rm cyl}$ 
our distances depend systematically as $D \propto T_{\rm cyl} \,\delta\kappa_0^{\rm em}/\cos i$.
The physical size (\FWHM{} in pc) has the same dependence.

The effect of doubling the emissivity on the relation between central column density and \FWHM{} is shown
in Fig.~\ref{fig_resultgauss}. The filaments have half the column density but are twice as extended. Wether the lower 
column density values are consistent with the model might be verifiable using extinction measurements.

For known $D$ and $T_{\rm cyl}$ the model provides a method to study the dust properties 
and to estimate the effective emission coefficient in the dense molecular phase.
The value is still uncertain because of the unknown inclination angle. 
On the other hand, the model would provide a method to estimate this angle
if $D$, $T_{\rm cyl}$, $\delta\kappa_0^{\rm em}$, and $\beta^{\rm em}$ were known.

\subsubsection{\label{sect_kappa}Dust properties}

The dust properties, the dust composition and sizes, in the dense molecular phase are still rather uncertain. 
Theoretically, the opacity could, depending on composition,
be a factor $\sim 10$ higher than in the diffuse phase \citep{Kruegel1994}. 
According to theoretical studies of grain coagulation the emission coefficient in the FIR/submm
regime should increase with density \citep{Ossenkopf1994}. 
Such a behavior is also indicated by recent studies of the dust emission spectrum for different column densities \citep{Martin2012}.
In this case it might be expected that the dust emission coefficient for cold filaments would increase as a function of $f_{\rm cyl}$ if the 
bounding pressure is approximately the same.
On the other hand, it needs to be considered that the effects of grain growth in dense clouds will be partly compensated
by stronger temperature variations which lead to a reduced {\it effective} emission coefficient. The effect 
can be disentangled by using a more realistic model for the dust temperature inside the filament. 

%



Measurements of the opacity $\delta \kappa^{\rm ext}$ in the submm regime
are generally based on correlations between the dust re-emission
and the column density of the gas assuming $\kappa^{\rm em}_\lambda=\kappa_\lambda^{\rm ext}$. 
The amount of gas and dust along the sight lines is derived using different
tracers depending on the phase considered. HI and CO observations are commonly used to trace the gas column 
density of the atomic and (up to a certain column density) molecular gas. 
Column densities in dense molecular gas
can also be based on the reddening of background stars. In most recent publications the SED
is modeled using a simple modified black body function assuming a single 
temperature as in Eq.~\ref{eq_emissivity}. Therefore, the derived opacity needs to be considered
as an approximate value. The correlation provides for given $\beta^{\rm ext}$ 
an estimate of the product $\delta \kappa_0^{\rm ext}$ or in units $\rm cm^2/H-atom$ the opacity
$\sigma_0^{\rm ext}=\delta\kappa^{\rm ext}_0 \,1.4 m_{\rm H}$.

Based on recent studies of the optically thin medium $\sigma_0^{\rm ext}$ lies in the range $(0.6 -1.6)\times 10^{-25}~{\rm cm^2/H-atom}$ \citep{Abergel2011a}. 
Our assumed value ($\sigma_0^{\rm em}=1.38\times 10^{-25}~{\rm cm^2/H-atom}$) lies at the high end of the derived range. In the dense molecular phase the effective emission coefficient
seems to be a few times higher than in the diffuse phase (see \citet{Martin2012} for a recent summary and discussion). 

For known $D$ and known $T_{\rm cyl}$ our physical model of self-gravitating filaments
provides an \emph{independent} estimate of the dust opacity.
As mentioned in Sect.~\ref{sect_observations} 
the previous distance estimates vary within a factor of two and seem to depend on the applied method.
The shortest distance of $D=460^{+40}_{-60}~{\rm pc}$
has been derived for the \object{Northern Streamer} using 
star counts \citep{Lada1994,Lada1999}, a value well in agreement with our results. 
Estimates based on the main sequence defined by 
B-type stars indicate for the cloud complex a distance around 1~kpc \citep{Harvey2008}. 
The model fit for filament a  together with the range of distances suggests for the
product $\delta \kappa_0^{\rm em}$ a physical range
\begin{equation}
	\sim 0.055~{\rm cm^2g^{-1}} \le\delta\kappa_0^{\rm em}\left(\frac{T_{\rm cyl}}{10~{\rm K}}\right)\frac{1}{\cos i}\le\sim 0.12~{\rm cm^2g^{-1}}
\end{equation}
where the lower value is related to the shortest distance $460~{\rm pc}$. As $T_{\rm cyl}$ in dense gas
probably has a physical lower limit around $10~{\rm K}$ \citepalias{Fischera2012a} the model indicates an upper limit 
$\delta \kappa_0^{\rm em} \sim 0.12 ~{\rm cm^2~g^{-1}}$ or $\sigma_0^{\rm em}\sim 2.8\times 10^{-25}~{\rm cm^2/H-atom}$.

\subsubsection{External pressure}


As we have shown in \citetalias{Fischera2012a}
the physical size (\FWHM{}) and the central column density $N_{\rm H}(0)$ 
of filaments of low to intermediate column density 
in \object{IC~5146}, \object{Polaris}, and \object{Aquila} as derived by \citet{Arzoumanian2011}
indicate $T_{\rm cyl} \sim 10~{\rm K}$ and a pressure range $\sim 1.5\times 10^4$ to $5\times 10^4~{\rm K ~cm^{-3}}$. 
For the same $T_{\rm cyl}$ and distance, we obtain similar values of $p_{\rm ext}$ here. However, it should be
remarked that the column densities in the aforementioned work were based on $\delta \kappa_0^{\rm em}=0.14~{\rm cm^2~g^{-1}}$
whereas in the modeling here we used $\delta\kappa_0^{\rm em}=0.0588~{\rm cm^2~g^{-1}}$.\footnote{In \citetalias{Fischera2012a} we 
corrected $N_{\rm H}(0)$ for the wrong $\bar \mu$ assumed in the work of \citet{Arzoumanian2011}.}

In a comparable study about the density structure of Bok-Globules  
\citet{Kandori2005} derived a range of external pressure values ranging from $2.1\times 10^4~{\rm K ~cm^{-3}}$ to
$1.8\times 10^5~{\rm K ~cm^{-3}}$. They agree within a factor of two with our study.

Inspection of Figs. \ref{fig_chisqr} and \ref{fig_chisqr2} reveals
the possibility of a common $p_{\rm ext}/k \sim 2\, \times 10^4~{\rm
K~cm^{-3}}$, with filaments `b' and `c' being on the high and low
side, respectively.  However, we feel that in the real ISM surrounding
these different filaments the pressure might actually vary: some filaments
might be embedded and therefore be surrounded by a medium with higher
pressure.  The model applied here and in other fields can then be used
to understand how uniform the pressure is in the ISM within a cloud
complex.


A factor two higher emission coefficient consistent with the larger distance 
would imply a considerably lower external pressure for IC 5146 ($p_{\rm ext}/k\sim 0.5\times 10^4~{\rm K~cm^{-3}}$).
As the larger distance also implies a large distance ($-61~{\rm pc}$) below the Galactic mid-plane (Sect.~\ref{sect_observations})
a lower pressure would be also expected considering the pressure profile of the ISM \citep{Ferriere2001}.

\subsection{\label{sect_lm}Luminosity to mass ratio}

%
Radiative transfer calculations provide a further method  to verify the assumption
made for the \emph{effective} dust emission behavior ($\kappa^{\rm em}_0$ and the exponent 
$\beta^{\rm em}$). Here, we used the $L/M$ ratio to validate the choice of $\kappa_0^{\rm em}$.

Relevant studies can be found in the work of \citet{Fischera2012b}. The estimates of the $L/M$ ratio 
are based on radiative transfer calculations through pressurized self-gravitating isothermal
cylinders and spheres which are illuminated by the interstellar radiation field assumed to be isotropic.
The calculations are based on dust properties close to the ones found in the diffuse interstellar
medium. The intrinsic mean properties are identical with the ones mentioned in Sect.~\ref{sect_emissivity}.

Considered are clouds with central extinction values in the range from $A_V=0.1~{\rm mag}$ up to $A_V=64~{\rm mag}$
for two different external pressures: $p_{\rm ext}/k=2\times 10^4~{\rm K/cm^3}$ and $p_{\rm ext}/k=10^6~{\rm K/cm^{3}}$.
The results for the high pressure regime do not seem to be of any relevance for the filaments analyzed in this work.

For cylinders with a central extinction $A_V=0$, 0.1, 1, 2, 4, 8, and 16~mag the corresponding $L/M$ ratios
are $1.15$, 1.051, 0.835, 0.685, 0.530, 0.396, and 0.200~$L_\odot/M_\odot$, respectively. As expected,
the ratio decreases towards filaments with higher column density. The radiation absorbed scales with the adopted 
interstellar radiation field.

For the assumed effective emission coefficient the derived column densities for filaments a, b, and c cover a 
range from $A_V\sim 5.6~{\rm mag}$ to $A_V\sim 12.1~{\rm mag}$ assuming, as in the radiative transfer calculations,
a total-to-selective extinction $R_V=3.1$ \citep{Fitzpatrick1999} and a gas-to-dust ratio
$N_{\rm H}/E(B-V)=5.8\times 10^{21}~{\rm cm^2~mag^{-1}}$ \citep{Bohlin1978}. The corresponding $L/M$ values
range from $0.363\pm 0.023$ to $0.180 \pm 0.025$ $L_\odot/M_{\odot}$. 
They are very close to the theoretical values of the radiative transfer calculations. This is reassuring, but
might well be fortuitous. The radiation field illuminating the filament as well as the absorption and scattering properties of 
the dust grains at UV and optical wavelengths might be different. In addition the extinction is uncertain by the unknown
inclination angle.



If we assume the larger cloud distance and therefore the higher effective emission coefficient
the column density is lower by a factor two. The
 $L/M$ ratio is (according to Eq.~\ref{eq_lummass}) increased by the same factor. 
For filament a for example $L/M=0.72\pm 0.04~L_\odot/M_\odot$ at $A_V\sim 2.8~{\rm mag}$.
In this case the observed ratio is $\sim 50\%$ higher than the corresponding theoretical value. 
However, this might be resolved given the uncertainties mentioned above.

\section{\label{sect_summary}Summary}

We have analyzed the structure of four compact filaments in the cloud complex \object{IC 5146} 
using recent \her\ observations taken with the SPIRE instrument which provides
the surface brightness at the reference wavelengths 250, 350, and $500~\mu{\rm m}$.
Scans of the surface brightness profiles were derived based on four aperture windows positioned
over parts of the filaments with well defined structure and background. 

For predicting the filament emission we applied the model of isothermal self-gravitating pressurized cylinders as described
in \citetalias{Fischera2012a}. In this model
the shape of the density profile is determined by the ratio $f_{\rm cyl}=(M/l)/(M/l)_{\rm max}$ of the actual and the
maximum possible mass line density. 
For simplicity we assumed the dust temperature to be constant throughout the filament.
The dust emissivity is approximated by a simple modified black-body spectrum with
an effective dust emission coefficient approximated through a power law 
$\kappa^{\rm em}_\lambda = \kappa^{\rm em}_0 (\nu/\nu_0)^{\beta^{\rm em}}$ with 
$\kappa^{\rm em}_0=10~{\rm cm^2 g^{-1}}$ at $250~\mu{\rm m}$
and $\beta^{\rm em}=1.8$. The dust-to-gas ratio in mass is assumed to be $\delta= 0.00588$. (The model depends
on the product $\delta\kappa_0^{\rm em}$).
The individual surface brightness profiles were modeled by taking the appropriate Gaussian approximations
of the psf into account.

Overall, we found good agreement between the modeled and the observed surface brightness profiles.
In accordance with the physical model filaments with higher central column density $N_{\rm H}(0)$ have
a smaller size (\FWHM{}) and show a steeper density profile.	
Based on the best example we derived a cloud distance $D\cos i = (493 \pm 36)
 (\delta\kappa_0^{\rm em}/0.0588~{\rm cm^2g^{-1}})(T_{\rm cyl}/10~{\rm K})~{\rm pc}$ and an external pressure
$(p_{\rm ext}/k)/\cos^2 i = 10^{4.35\pm 0.13}(\delta\kappa_0^{\rm em}/0.0588~{\rm cm^2g^{-1})^{-2}}~{\rm K~cm^{-3}}$,
where $i$ is the projection angle of the filament with $i=0$ for an edge-on view.
The previously estimated distances of the cloud complex limit $\delta \kappa_0^{\rm em}$ to the range
$\sim 0.055~{\rm cm^2g^{-1}}\le\delta\kappa_0^{\rm em}(T_{\rm cyl}/10~{\rm K})/\cos i\le\sim 0.12~{\rm cm^2g^{-1}} $.
An upper limit is therefore $\delta\kappa_0^{\rm em} \sim 0.12~ {\rm cm^2g^{-1}}$.

Considering the simplifications the results are encouraging 
for further investigations. They should be based on a better resolution of the filaments (more nearby clouds), a 
complete coverage of the SED (including PACS), and a more realistic model of the
temperature variation in the filament. This will 
provide a better estimate of the mass ratio $f_{\rm cyl}$, the external pressure and the distance, and
possibly allow investigation of the intrinsic dust properties.  It might also
probe to what extent  the dust properties can be considered constant within a single dense filament.

\acknowledgement{This work was supported by grants from the Natural
  Sciences and Engineering Research Council of Canada and the Canadian
  Space Agency. J. Fischera would like to thank Dr. R. Tuffs for helpful comments.
  We like to thank the referee and the editor for critical comments which 
  helped to improve the manuscript.}

\appendix

\section{\label{app_surfbrightness}Predicting surface brightnesses}

The filament is located in a distance where the resolution affects the observed profiles. To take this effect
into account we used the Gaussian approximation of the $psf$ of the SPIRE instrument.
The theoretical surface brightness profile from the cylinder measured at a pixel resolution $\Delta \vartheta$ is then given by
\begin{equation}
	\hat I^{\rm fil}_{\nu_{r}}(\vartheta_i) = \frac{1}{\Delta\vartheta_i}\int\limits_{\vartheta_i-\frac{1}{2}\Delta \vartheta_i}^{\vartheta_i+\frac{1}{2}\Delta \vartheta_i}{\rm d}\vartheta\,\int{\rm d}\vartheta'\,psf_{\nu_{r}}(\vartheta-\vartheta')\,\bar I^{\rm fil}_{\nu_{r}}(\vartheta')
\end{equation}
where $\bar I_{\nu_{r}}$ is the theoretical monochromatic flux of the broad-band filter at reference frequency $\nu_{r}$.
In our calculations we made the convolution at 10 time higher resolution than the pixel size of the scan.

As the $psf$ varies along the filter with wavelength and depends therefore on the dust emission spectrum it would
be more accurate to derive the monochromatic flux for each filter after the convolutions. Here, we consider the 
effect to be small. A larger uncertainty might be related to use the Gaussian approximation instead of the actual $psf$.
We note that our approach here is based on the assumption that the dust emission spectrum does not vary
as a function of the impact parameter. This is certainly not true in the case of varying dust temperature that could 
exist because of radiative transfer effects which produce a declining dust temperature from the edge to the cloud centre
or because of a non isotropic radiation field. To provide the most accurate estimates of the surface brightness profiles 
one should derive the actual convolved profiles $I^{\rm fil}_\nu(\vartheta_j)$ for a number of frequencies $\nu$ which
allow a determination of the monochromatic broad band fluxes.

The parameters of the filaments were derived using a non-linear $\chi^2$-fit where
the partial derivatives for the parameters $a_j$ were estimated in accordance with
\begin{eqnarray}
	\frac{\partial \hat I^{\rm fil}_{\nu_{r}}}{\partial a_j}(\vartheta_i)&=& \frac{1}{\Delta\vartheta_i}\int\limits_{\vartheta_i-\frac{1}{2}\Delta \vartheta_i}^{\vartheta_i+\frac{1}{2}\Delta \vartheta_i}{\rm d}\vartheta \nonumber\\
	&& \times \int{\rm d}\vartheta'\,psf_{\nu_{r}}(\vartheta-\vartheta')
		\frac{\partial \bar I^{\rm fil}_{\nu_{r}}(\vartheta')}{\partial a_j}.
\end{eqnarray}

\section{\label{app_source}Source subtraction from the scan}

The structure of the ISM makes it difficult to find filaments on a flat or well defined background. 
We want to consider an aperture window containing, as in case of the aperture window chosen for filament c,
not only the emission of the filament and a smoothly varying background
emission but also of a marginally extended source.

We modeled the source as an elliptical Gaussian function convolved with
the $psf$ also assumed to be Gaussian.
The major or minor axis is further assumed to be aligned with the scan direction. 
The surface brightness from the source is therefore assumed to be
\begin{equation}
	S_{\nu}(x,y) = \frac{F_{\nu}}{2\pi \sigma_x\sigma_y} \exp\left\{-\frac{(x-x_0)^2}{2\sigma^2_x}
		-\frac{(y-y_0)^2}{2\sigma^2_y}\right\}
\end{equation}
where $F_\nu$ is the flux density given by
\begin{equation}
	F_{\nu} = \frac{\delta M_{\rm cl}}{D^2}\kappa^{\rm em}_0(\nu/\nu_0)^{\beta} B_{\nu}(T_{\rm dust})
\end{equation}
where $\delta\kappa^{\rm em}_0=0.0588~{\rm cm^2 g^{-1}}$ and $\beta=1.8$ and where $\nu_0=c/\lambda_0$ where $\lambda_0=250~\mu{\rm m}$.
The standard deviations of the extra emission along the window width and the window length ($x$ and $y$) are given by
\begin{equation}
	\sigma_x(\nu) = \sqrt{\sigma_S^2+\sigma_{\rm psf}^2(\nu)},\quad \sigma_y(\nu)=\sqrt{\epsilon^2\sigma_S^2+\sigma^2_{\rm psf}(\nu)}
\end{equation}
where $\sigma_S$ and $\epsilon\sigma_S$ are the standard deviations of the source vertical and parallel to the 
scan direction and where $\sigma_{\rm psf}$ is the standard deviation of the $psf$.
The source emission is described by the dust temperature $T_{\rm dust}$, the cloud mass $M_{\rm cl}$, the source size $\sigma_S$, and the source aspect ratio 
$\epsilon$.

To avoid any additional systematic uncertainties we made no further assumption for the surface
brightness contributions from the background and the filament other than assuming that the surface brightness is constant along
the scan width and only varies along the scan length. The surface brightness along the scan length is
given by the mean
\begin{equation}
	\left<I_{\nu_{r}}\right>(y_j) = \frac{1}{N_x}\sum\limits_{i=1}^{N_x} I_{\nu_{r}}(x_i,y_j)-\bar S_{\nu_{r}}(x_i,y_j).
\end{equation}
where $\bar S_{\nu_{r}}$ is the monochromatic flux for the filter with reference frequency $\nu_{r}$ and where
$N_x$ is the number of pixel along the aperture width.

The source parameters were derived from the residual
\begin{equation}
	\Delta I_{\nu_{r}}(x_i,y_i) = I_{\nu_{r}}(x_i,y_j) - \left<I_{\nu_{r}}\right>(y_j)
\end{equation}
by minimizing iteratively
\begin{equation}
	\chi^2 = \sum\limits_{r=1}^{3} \sum_{i,j}\frac{(\bar S_{\nu_{r}}(x_i,y_j)-\Delta I_{\nu_{r}}(x_i,y_j))^2}{\sigma^2_{r}}.
\end{equation}

\bibliographystyle{aa} 
\bibliography{fischerareference}

\end{document}